\documentclass{aa}  

\usepackage{graphicx}
\usepackage{color}
\usepackage{hyperref}
\begin{document}

\title{A determination of the Large Magellanic Cloud dark matter subhalo mass using the Milky Way halo stars in its gravitational wake}
\titlerunning{Large Magellanic Cloud dark matter subhalo mass determination}

\author{K. J. Fushimi
          \inst{1} 
          \and
           M. E. Mosquera \inst{1}$^,$\inst{2} \and  M. Dominguez\inst{3}}

   \institute{Facultad de Ciencias Astronómicas y Geofísicas, University of La Plata. Paseo del Bosque S/N 1900, La Plata, Argentina.\\       \email{kfushimi@fcaglp.unlp.edu.ar}\\
         \and
             Dept. of Physics, University of La Plata, c.c.~67 1900, La Plata, Argentina.\\       
         \and 
             Instituto de Astronom\'{\i}a Te\'orica y Experimental, (IATE-CONICET), Observatorio Astron\'omico de C\'ordoba, Universidad Nacional de C\'ordoba, Laprida 854, X5000BGR, C\'ordoba, Argentina.CONICET, CCT C\'ordoba \\    
             }

   \date{}

\abstract{}{Our goal is to study the gravitational effects caused by the passage of the Large Magellanic Cloud (LMC) in its orbit on the stellar halo of the Milky Way.}{We employed \textit{Gaia} Data Release 3 to construct a halo tracers dataset consisting of K-giant stars and RR-Lyrae variables. Additionally, we compared the data with a theoretical model to estimate the dark matter subhalo mass.}{We have improved the characterisation of the local wake and the collective response due to the LMC's orbit. We have also estimated for the first time the dark subhalo mass of the LMC to be of the order of $1.7\times 10^{11}$ M$_{\odot}$, which is comparable to previously reported values in the literature.}{}
 
   
   
   
   


   \maketitle
%

\section{Introduction}
Dark matter (DM) is central to the standard cosmological model (LCDM), providing gravitational support for the formation of galaxies and systems of galaxies \citep{mo_vandenbosch_white_2010}. Its existence is backed by a plethora of observational data, including galaxies' rotation curves \citep{Zwicky:1933,Rubin:1970}, strong and weak gravitational lensing effects \citep{Massey:2010,Clowe:2006}, and even the presence of baryonic acoustic oscillation \citep{Planck18} in the earlier gravitational wells observed in the cosmic microwave background (CMB).

Despite these successes, we still lack a precise detection either in the laboratory \citep{ Bernabei:2022,Aprile:2023,Barberio:2022,ANAIS:2022} or by indirect astrophysical observation \citep{HEES:2022,VERITAS:2023,MAGIC:2023}.  Many theoretical candidates arise out of physics beyond the standard model (BSMP), like weakly interacting massive particles (WIMPS). These massive ($m_{DM} \sim 100$ GeV) particles could weakly interact with nucleons, and therefore their signals have been looked at by several laboratory and accelerator experiments. Also, their annihilation signals could be detectable through $\gamma$-ray emission by high-energy telescopes. Despite a massive experimental effort,  DM remains a theoretical hypothesis, albeit one with impressive empirical support.

Other DM candidates could be more massive, such as primordial black holes (PBHs) \citep{doi:10.1146/annurev-nucl-050520-125911}, which were recently constrained with a series of consistency tests. Nowadays, there is still room to be an essential contributor to the DM content but this is limited to small windows in mass \citep{10.3389/fspas.2021.681084}. Other candidates include massive ultralight particles (ULDMs) that could reach masses as low as $10^{-23}$ eV \citep{PhysRevD.95.043541}. Therefore, the possible DM mass range remains unconstrained today. Additionally, DM particles could interact with themselves, and have or not have spin, and other properties, including their mass, remain elusive.

Depending on the nature of the DM particle, there are relevant changes in the structure and number of DM halos and subhalos \citep{galaxies7040081}. For example, some candidates, like warm dark matter (WDM), introduce a cut-off scale in the initial power spectrum of mass fluctuations ($m_{DM} \sim 1$ KeV), and others a scale during the non-linear evolution phase in which the DM particles self-interact (SIDM) \citep{TULIN20181}. Both processes change the abundance of DM subhalos and the density profiles of the DM halos in comparison to the predictions of the cold dark matter (CDM) model.

\citet{Buschmann:2018} show, using CDM simulations, that the gravitational pull of DM subhalos affects the distribution of stars in galactic halos, and could be used to discover dark subhalos (those without star formation in situ, a precise prediction of the CDM model) and also allow one to test the nature of the DM particles itself. This work simulated a passing DM subhalo's perturbation to the phase space stellar distribution. Stars are pulled towards the subhalo as it passes, leaving a distinctive feature in the halo stars' velocity and number density, known as a wake. This phenomenon was previously analytically described by \citet{1986ApJ...300...93W} due to the gravitational friction that provokes the orbital decay of the satellite galaxies that inhabit the DM subhalo. There have been some efforts to quantify the magnitude of phase-space perturbations caused by the passage of DM subhalos using simulations and their possible detection \citep{Bazarov:2022}.

\citet{Garavito:2019} used CDM simulations to quantify the impact of the LMC's passage on the density and kinematics of the Milky Way (MW) DM halo and the observability of these structures in the MW's stellar halo. Their results indicated a pronounced wake, which could be decomposed in a transient response and a collective one in both the DM and stellar halo distributions. Such an effect was observed for the first time in the MW halo stars by \citet{Conroy:2021}. The authors studied the effects induced by the Magellanic Clouds System (MCS) merging in selected samples of MW stars with precise \textit{Gaia} Satellite measurements. This detection, and the increasing availability of stellar data from \textit{Gaia} DR3 release, has paved the way for a more precise measurement of the effect. Measurements of the wake on the perturber systems of reference will allow pursuit further testing of the DM particle's nature using this merger data, as was proposed recently by \citet{Foote:2023} (in the context of systems of galaxies see \citet{Furlanetto:2002,Buehler:2023}).  Furthermore,  \citet{Aguilar:2022} and  \citet{Cunningham:2020} conducted studies on the decomposition into spherical harmonics of both density and velocity, respectively, in order to quantify the response of the DM halo to the passage of the Large Magellanic Cloud (LMC).

Based on the findings of \citet{Conroy:2021}, we have used \textit{Gaia} Data Release 3 to study the DM subhalo of the Magellanic Clouds. Our code developments will also allow us to apply the methodology to other MW satellite galaxies and globular clusters and even to develop methods of detecting the presence of the dark subhalos predicted by the CDM model. 

This work is organised as follows. In Section \ref{data}, we present the data samples selection methodology used to identify the effects of the DM subhalo of the LMC on the MW stellar halo. In Section \ref{like}, we briefly describe the theoretical model. Meanwhile, we present our results in Section \ref{resultados}. Section \ref{conclusiones} presents the conclusions and future perspectives. Finally, we present in Appendix \ref{apendice} a description of the coordinates' transformation, in Appendix \ref{apendice:machinelearning} the machine learning algorithm used to estimate radial velocities, and in Appendix \ref{apendice:distancias} a validation of our method of inferring distances.


\section{Data reduction}
\label{data}
We studied the gravitational response of the MW's halo to the passage of the LMC in its orbit. To achieve this, we used the data from \textit{Gaia} Data Release 3 \citep{GAIA:DR3,Gaia:2016} and created two catalogues of halo tracers; namely, K-giant stars and RR-Lyrae ones. We followed the steps proposed by \citet{Conroy:2021} to address this task.
\subsection{K-giant dataset}
To construct the K-giant catalogue, we started the analysis with 162240774 sources characterised by $ruwe$ values below 1.4, parallax measurements lower than 0.2 mas, and galactic latitudes of $|b|>10^{\circ}$ to remove the galactic plane. To ensure data quality, we performed a series of cleaning procedures. First, we eliminated sources lacking proper motion and photometric data.  To account for dust extinction, we obtained the dust map from \citet{dustmap} and considered the Schlegel, Finkbeiner $\&$ Davis (SFD) map to derive the excess colour, $E(B-V)$. Subsequently, we discarded all sources with $E(B-V)>0.3$. To obtain the corrected magnitudes, we considered the following coefficients: $A_G/E(B-V) = 2.4$, $A_{BP}/E(B-V) = 2.58$, and $A_{RP}/E(B-V) = 1.65$. To focus solely on the giant branch, we restricted the selection to sources satisfying the condition $1.4 < (BP^* - RP^*) < 2$, where $BP^*$ and $RP^*$ represent the corrected magnitudes. Next, following \citet{Riello:2021}, we performed the $3\sigma$ cut upon the corrected BP and RP flux excess factor ($C^*$). After completing the data-cleaning process, to ensure the purity of our catalogue specifically for K-giant stars, we performed a cross-match with the spectral types provided by \textit{Gaia} (gaiadr3.astrophysical parameters) and obtained a dataset of 490669 sources. Finally, we restricted our analysis to objects within a galactocentric distance between $30$ kpc and $100$ kpc, leaving 245086 sources. Among them, only 10989 had radial velocities measured by \textit{Gaia} \citep{Katz:2023}.

To estimate the radial velocity for the remaining sources, we employed a machine learning algorithm, specifically a RandomForestRegressor \citep{scikit-learn}. The accuracy of our model is 87.0$\%$ (see Appendix \ref{apendice:machinelearning} for details).

To determine the photometric distance, we used the MIST code \citep{MESA_1,MESA_2,MESA_3} to generate an isochrone with the specific LMC parameters, which are an age of $10$ Gyr and a metallicity of $\textrm{[Fe/H]} = -1.5$. We restricted the isochrone to an effective temperature from $3800$ K to $4400$ K, and fitted the polynomial equation 
\begin{equation}
\label{mg}
 M_G = 2.8894(BP^{*}-RP^{*})^2-11.9263(BP^{*}-RP^{*})+8.7151\,.   
\end{equation}

To validate our distance inference method, we compared our calculated distances with reference values for some globular clusters, the Magellanic Clouds (LMC and SMC), and some satellite galaxies (see Appendix \ref{apendice:distancias} for details). Our method successfully reproduced tabulated distances, since the mean calculated distances differed by less than $20\%$.

Afterwards, we implemented several masks to exclude known objects from our analysis. Specifically, we applied angular and distance masks to all the globular clusters listed in \citet{Harris:1996} and all the satellite galaxies reported in \citet{Drlica-Wagner:2020}.

Following the methodology proposed by \citet{Conroy:2021}, we employed proper motions to eliminate structures linked to the Sagittarius stream. To achieve this, we initially correct the proper motions due to the solar reflex motion, with the Gala package \citep{gala,gala2}.  The used parameters were $R_{\odot} = 8.122$ kpc \citep{GRAVITY}, $(V_{R,\odot}, V_{\phi,\odot}, V_{Z,\odot}) = (-12.9, 245.6, 7.78)$ km/s \citep{Drimmel}, and the distance of the Sun from the Galactic mid-plane, $Z_{\odot} = 20.8$ pc \citep{Bennett}. For $b > 10^{\circ}$ and $|B_{Sgr}| < 15^{\circ}$, where $B_{Sgr}$ is the latitude in the frame of the Sagittarius orbital plane, we removed part of the northern arm of the stream by taking out the sources that have $\mu_{\alpha*}> -1.3$ mas/yr, $-0.4 < \mu_{\delta}< 0.3$ mas/yr, and $\mu_{\delta} > 1.7\mu_{\alpha*} + 0.4$. To eliminate the rest of the north arm, we applied a mask to the region with coordinates $b>0^{\circ}$ and $180^{\circ}<l<210^{\circ}$. The final selection of sources was based on proper motions; that is, we kept only those that satisfied $\mu_{\alpha*}^2+(\mu_{\delta}+0.1)^2<0.5^2$ \citep{Conroy:2021}. By implementing this criterion, one effectively excludes disk stars, stars belonging to the Large and Small Magellanic Clouds, the Sagittarius dwarf spheroidal, and other Sagittarius arms. After this matching, our final dataset of K giants had 6058 sources.
\subsection{RR-Lyrae dataset}
To build the RR-Lyrae catalogue, we started the process by using the 271779 sources catalogued as RR-Lyrae variables by \textit{Gaia}. Initially, we performed data cleaning by keeping stars with $ruwe<1.4$, and excluding those lacking metallicity, as well as those with errors in metallicity exceeding the absolute value of the metallicity itself. We then discarded all sources with $E(B-V)>0.3$, and the galactic plane $|b|<10^{\circ}$. These cuts yielded a set of  66610 stars, of which only 2440 had radial velocity measurements provided by \textit{Gaia}. 
 
To increase the amount of data with measured radial velocities and metallicities, we performed a series of data cross-matching steps. We used the SEGUE catalogue \citep{sloan:2012} to complete the radial velocities and metallicities of our dataset, using the ones measured by Sloan (cross-match with ID table {\it sdss$\_$dr17.x1p5$\_\_$specobj$\_\_$gaia$\_$dr3$\_\_$gaia$\_$source}). In this case, no extra points were incorporated.

Additionally, we utilised the RR-Lyrae catalogue provided by \citet{wang:2022} to complete our catalogue with the sources not present in our dataset (288 points were incorporated) or to complete our radial velocity and metallicity data. After the cross-matching process, we ended up with a catalogue of 73598 sources, of which 6523 have measured radial velocities. To obtain the radial velocity for the remaining sources, we applied a combined algorithm of data augmentation + random-forest regressor.   
In this case, the accuracy of our model is 49.0$\%$ (see Appendix \ref{apendice:machinelearning} for details). Similar to the K-giant approach,  we corrected the magnitudes to account for dust extinction. The absolute magnitude is connected to the metallicity through $M_G = 0.32\textrm{[Fe/H]} + 1.11 $ \citep{Muraveva:2018}; therefore, one can obtain their distances using the distance modulus relationship.

To ensure consistency, we followed a similar approach to that used with the K-giants. To eliminate known objects from the \citet{Harris:1996} and \citet{Drlica-Wagner:2020} catalogues, we applied an angular mask, taking into account the distance to the sources. Additionally, to exclude the Sagittarius stream from our analysis, we employed the cut-off criterion of $|B_{Sgr}|<15^{\circ}$ for $b>0^{\circ}$. Then, we focused our analysis on objects with galactic distances between $30 \textrm{ kpc} < R_{gal}< 100 \textrm{ kpc}$. In the final step, we removed stars that did not satisfy the condition $\mu_{\alpha*}^2+(\mu_{\delta}+0.1)^2<0.5^2$ \citep{Conroy:2021} after performing the solar reflex motion correction to the proper motions. Therefore, the final sample has 2446 sources.

Since our aim is to extract the mass of the DM subhalo surrounding the Large and Small Magellanic Clouds, we performed a transformation of coordinates to a new reference system. This particular coordinate reference system is centred on the centre of mass (CM) of the MCS, with the $x$ axis aligned with the direction of the velocity of the CM (see appendix \ref{apendice} for details), and it is considered a rest frame. In order to obtain the position and velocity of the CM, we considered the LMC mass to be nine times the SMC mass \citep{Craig:2021}.
\section{Theoretical model and likelihood analysis}

\label{like}
We used the theoretical model for stellar wakes from DM subhalos proposed by \citet{Buschmann:2018}, where the authors assumed a Plummer sphere for the density profile of the DM subhalo. The reference system used in this section corresponds to the one centred on the CM of the MCS, with the $x$ axis in the direction of the velocity of the DM subhalo mass (see Appendix \ref{apendice-cm} for details). From the collisionless Boltzmann equation, they derived the time-independent phase-space distribution function in the subhalo rest frame
\begin{align}
f\left(\bar{r},\bar{v},M_s\right)&=f_0(\bar{v}) \left(1+ \frac{2G M_s}{v_0^2} \left(\bar{v}+\bar{v}_s\right) \cdot \bar{\alpha}\right) \label{f1-1} \,, \\
\bar{\alpha}\left(\bar{r},\bar{v},M_s\right)&=\frac{1}{r v \sqrt{1+\frac{R_s^2}{r^2}}} \frac{\sqrt{1+\frac{R_s^2}{r^2}} \frac{\bar{v}}{v}-\frac{\bar{r}}{r}}{\sqrt{1+\frac{R_s^2}{r^2}}-\frac{\bar{v}\cdot \bar{r}}{r v}} \,.\nonumber 
\end{align}

In the equations, $f_0(\bar{v})=\frac{n_0 e^{-\left(\bar{v}+\bar{v}_s\right)^2/v_0^2}}{\pi^{3/2} v_0^3}$, ${v_0= \sqrt{2} \sigma_v}$ where $\sigma_v$ is the velocity dispersion, $\bar{v}_s$ and $M_s$ are the DM subhalo velocity and mass, respectively, ${R_s(M_s)=1.62 \sqrt{M_s/(10^8 \rm{M}_{\odot})}}$, and $n_0$ is the star density inside the region of interest. This expression can be easily extended for three different velocity dispersions by including $v_{0x}$ $\left(\sigma_{v_x}\right)$, $v_{0y}$ $\left(\sigma_{v_y}\right)$, and $v_{0z}$ $\left(\sigma_{v_z}\right)$. In this case, the distribution function in the subhalo rest frame can be written as
\begin{align}
f\left(\bar{r},\bar{v},M_s\right)&=\frac{n_0 \,e^{-\left(\frac{v_x+v_s}{v_{0x}}\right)^2-\left(\frac{v_y}{v_{0y}}\right)^2-\left(\frac{v_z}{v_{0z}}\right)^2}}{\pi^{3/2} v_{0x} v_{0y} v_{0z}} \left(1+2G M_s \beta \right) \label{f1-3}\,, \\
\beta \left(\bar{r},\bar{v},M_s\right)&= \left(\frac{v_x+v_s}{v_{0x}^2} \hat{i} +\frac{v_y}{v_{0y}^2} \hat{j} +\frac{v_z}{v_{0z}^2} \hat{k} \right) \cdot \bar{\alpha}\,. \nonumber
\end{align}

In order to obtain the mass of the DM subhalo of the MCS, we performed a statistical analysis using the likelihood function to compare observational data in the new reference system and the theoretical model. This analysis was performed by using only the space data ($3D$) and the phase-space data ($6D$). The un-binned likelihood function for the $6D$ analysis is \citep{Buschmann:2018}
\begin{equation}
    p_{6D}(M_s,\theta)= e^{-N_s(M_s)} \prod_{k=1}^{N_{d}} f\left(\bar{r}_k,\bar{v}_k,M_s\right)\,, \label{p6}
\end{equation}
where $N_{d}$ is the number of stars in the region of interest (sphere of radius $R$ centred on the CM of the MCS), $N_s$ is the predicted number of stars in the same region, and $\theta$ are the fixed parameters of our model; that is, $n_0$, $v_0$, $R_s$, and $\bar{v}_s$. For the Plummer sphere and for the distribution function of Eq. (\ref{f1-1}), $N_s(M_s)$ can be computed as \citep{Buschmann:2018}
\begin{align}
N_s(M_s)&=\frac{4}{3}\pi R^3 n_0 +\frac{4 \pi G M_s n_0}{v_0 v_s}\, \gamma \,F\left(\frac{v_s}{v_0}\right) \,, \nonumber\\
\gamma &=R^2\sqrt{1+\frac{R_s^2}{R^2}} - R_s^2\, {\rm arcsinh} \left(\frac{R}{R_s}\right) \,,\nonumber \\
F(x)&= e^{-x^2} \int_0^x e^{y^2} {\rm d}y \,.\nonumber 
\end{align}

For the Plummer sphere and for the distribution function of Eq. (\ref{f1-3}), the predicted number of stars is
\begin{align}
N_s(M_s)&= \frac{4}{3}\pi R^3 n_0  +\frac{4 G M_s n_0}{v_{0x} v_{0y} v_{0z}} \,\gamma \, I\left(\bar{v}_0,v_s\right)\,,\nonumber \\
\begin{split}
I\left(\bar{v}_0,v_s\right)&=\int \frac{{\rm d}^3 p}{2\pi} \, e^{-p^2} \cos\left(\frac{2 v_s  p_x}{v_{0x}}\right) \\
&\hspace{0.7cm}\left(\left(\frac{p_x}{v_{0x}}\right)^2+\left(\frac{p_y}{v_{0y}}\right)^2+\left(\frac{p_z}{v_{0z}}\right)^2\right)^{-1/2} \,.
\end{split}\nonumber
\end{align}

For the $3D$ data, the un-binned likelihood function is
\begin{align}
p_{3D}(M_s,\theta)&= e^{-N_s(M_s)} \prod_{k=1}^{N_{d}} \int {\rm d}^3 v \, f\left(\bar{r}_k,\bar{v},M_s\right)\,. \label{p3}
\end{align}

To determine the DM subhalo mass, we used a Markov-Chain Monte Carlo (MCMC) method. For this purpose, we utilised the emcee package \citep{emcee}. We considered the function
\begin{align}
\lambda(M_s)&=\ln\left(p_{x}\left(M_s, \theta\right)\right) \, \, ,
\end{align}
where $x$ stands for the 3D or 6D analysis. The prior used was $10<\log_{10} M_s<12$. The functions, $p_x$, are presented in Eqs. \eqref{p6} and \eqref{p3}. We considered $32$ walkers and checked the convergence every $100$ steps. To compute the uncertainties, we used the $16^{th}$, $50^{th}$, and $84^{th}$ percentiles of the samples in the marginalised distributions \citep{emcee}.

\section{Results}
\label{resultados}
\begin{figure*}
  \sidecaption
  \includegraphics[width=12cm]{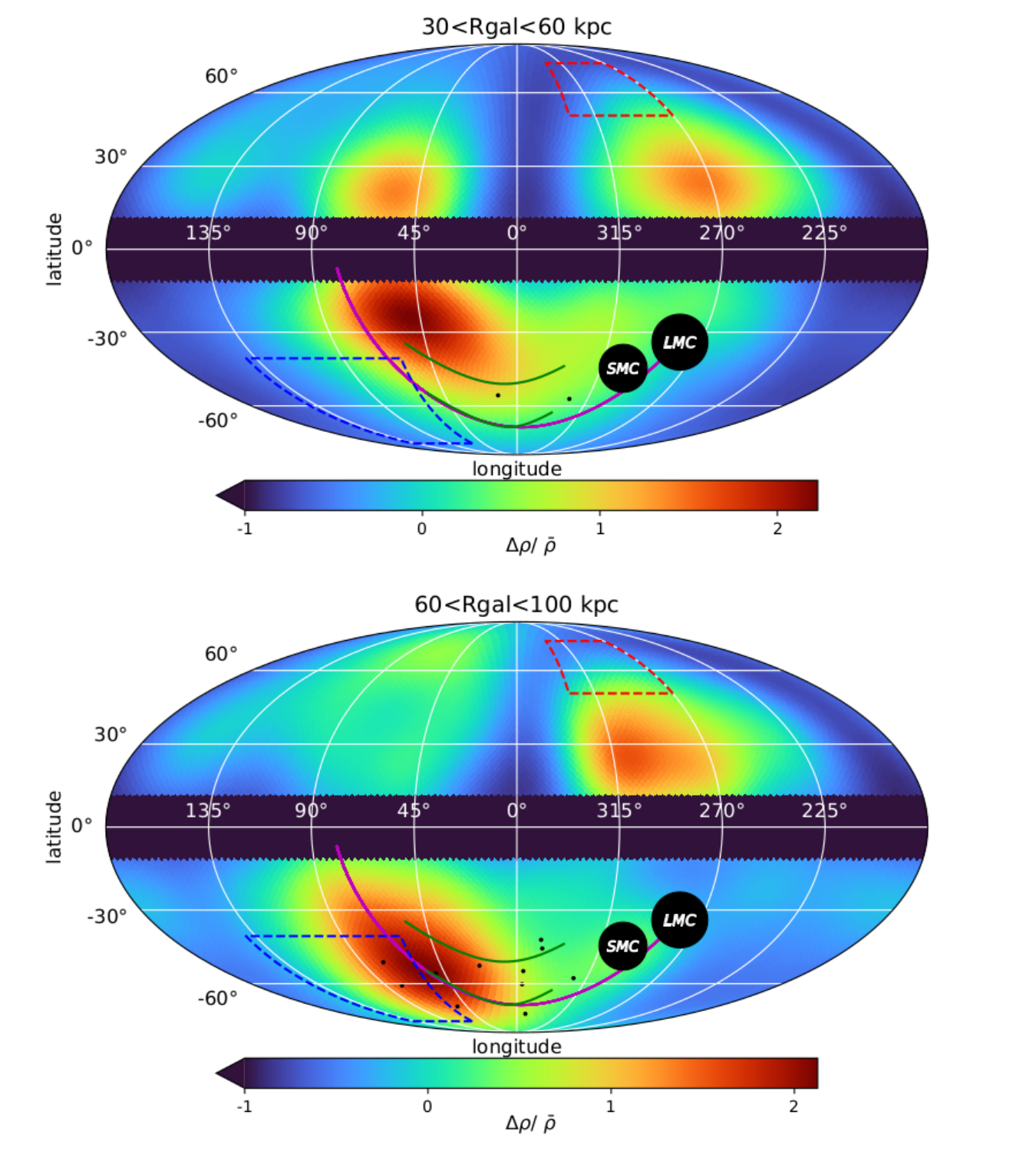}
      \caption{ Density distribution of K-giant and RR-Lyrae variables (Mollweide projection map) with $30<R_{gal}<60 $ kpc (top panel) and $60<R_{gal}<100 $ kpc (bottom panel). The data was smoothed using an FWHM of $30^\circ$. The magenta line represents the past orbit of the MCS CM. The dashed blue and red lines represent the Pisces and Virgo overdensities, respectively. The black dots represent the members of the Magellanic Stellar Stream. The green lines are the Pisces Plume.}
         \label{mollview}
   \end{figure*}

 In Fig. \ref{mollview}, we present a Mollweide projection map displaying the distribution of our final dataset of 8504 stars in galactic coordinates (6058 K giants and 2446 RR Lyraes). To enhance the visual representation, the map has been smoothed using Gaussian functions with a full width at half maximum (FWHM) of $30^{\circ}$. The colour bar represents the density contrast, indicating the relative density variation from its mean value across the sample. The past 1 Gyr orbit of the CM of the MCS is shown with a magenta line, computed with the gala package using the MilkyWayPotential \citep{Bovy:2015}. The dotted blue and red lines represent the Pisces \citep{Chandra:2023b} and Virgo overdensities \citep{Perottoni:2022}, respectively. The green lines are the polynomials adjusted by \citet{Chandra:2023b} that limit the Pisces Plume. The black dots represent the members of the Magellanic Stellar Stream taken from \citet{Chandra:2023a}. The dark blue region corresponds to the masked region representing the galactic plane.
   
Two distinct regions of overdensities can be observed. The first one, located in the northern hemisphere, with a longitude range between 225° and 315°, is associated with the collective response, also known as the global response. On the other hand, the southern feature appears to cover a larger area  ($-30^{\circ}<l<130^{\circ}$) and exhibits significant prominence at a longitude of $50^{\circ}$ and a latitude of approximately $-26^{\circ}$. This overdensity is associated with the local wake. A more minor component is also present in the northern hemisphere, within a longitude range of 30° to 90°. It appears separate from the southern component due to the masking of the galactic plane, implemented to prevent contamination. The intensity of the wake is greater than that of the collective response.  The ratio between the counts per pixel of the wake at $l=50^{\circ}$, $b=-26^{\circ}$ and the counts per pixel of the collective response at $l=279^{\circ}$, $b=24^{\circ}$ (the coordinates of the highest overdensity of the collective response) is 1.29, considering the complete dataset. As one can see, the CM of the MCS past orbit is located over the local wake, and the deviations could be signalling the effect of the DM mass of the wake according to the results of \cite{Foote:2023}.

In the upper panel of Fig. \ref{mollview}, an inner region of the halo is shown (sources between 30 and 60 kpc from the galactic centre). It can be noted that both the wake and the collective do not correspond to either the Pisces or the Virgo overdensities. In previous works \citep{Belokurov:2019,Conroy:2021}, it was noticed that there exists a subregion produced by the Magellanic Clouds called the Pisces Plume. The southern overdensity that we identified as the wake does not fall within this region. 

On the other hand, in the lower panel, we plot the outer region of the halo, between 60 and 100 kpc, a region studied by \citet{Belokurov:2019} and \citet{Conroy:2021}. Once again, the collective does not belong to the Virgo overdensity. Nevertheless, the global response could be truncated due to the masking of the galactic plane and the Sagittarius stream. However, part of the wake lies on the edge of the Pisces overdensity, and the maximum of the wake is indeed located in the Pisces Plume. Additionally, it has been verified that none of the catalogued points belonging to the Magellanic Stellar Stream \citep{Chandra:2023a} are found in our dataset.

Comparing our results with the ones obtained by \citet{Conroy:2021}, we observe some slight differences in the locations of the overdensities. Specifically, our sample's maximum southern overdensity is slightly displaced further north. Similarly, the maximum northern overdensity in our sample is slightly shifted towards the south and east. In particular, we have also compared only our K-giant sample results with the final public catalogue developed by \citet{Conroy:2021}. The coordinates of the maximum overdensities of the local wake and the collective response for our dataset are ($l=52^{\circ}$, $b=-24^{\circ}$) and ($l=275^{\circ}$, $b=25^{\circ}$), respectively; meanwhile, for Conroy's data they are approximately ($l=49^{\circ}$, $b=-54^{\circ}$) for the wake and ($l=326^{\circ}$, $b=54^{\circ}$) for the collective. Our definition of the local wake is larger than the one proposed by \citet{Conroy:2021}, and the ratio between the counts per pixel of the wake and the counts per pixel of the collective response at each maximum is 1.5 for our data and  1.33 for Conroy's data-set. However, if we consider only K giants located within $60<R_{gal}<100$ kpc, we successfully replicated the wake's position, with its peak occurring at $l=57^{\circ}$, $b=-51^{\circ}$. Moreover, when we compare our map with the simulations presented in \citet{Conroy:2021}, we observe quite an agreement regarding the positions of the overdensities.

\begin{figure}
    \centering
   \includegraphics[width=\hsize]{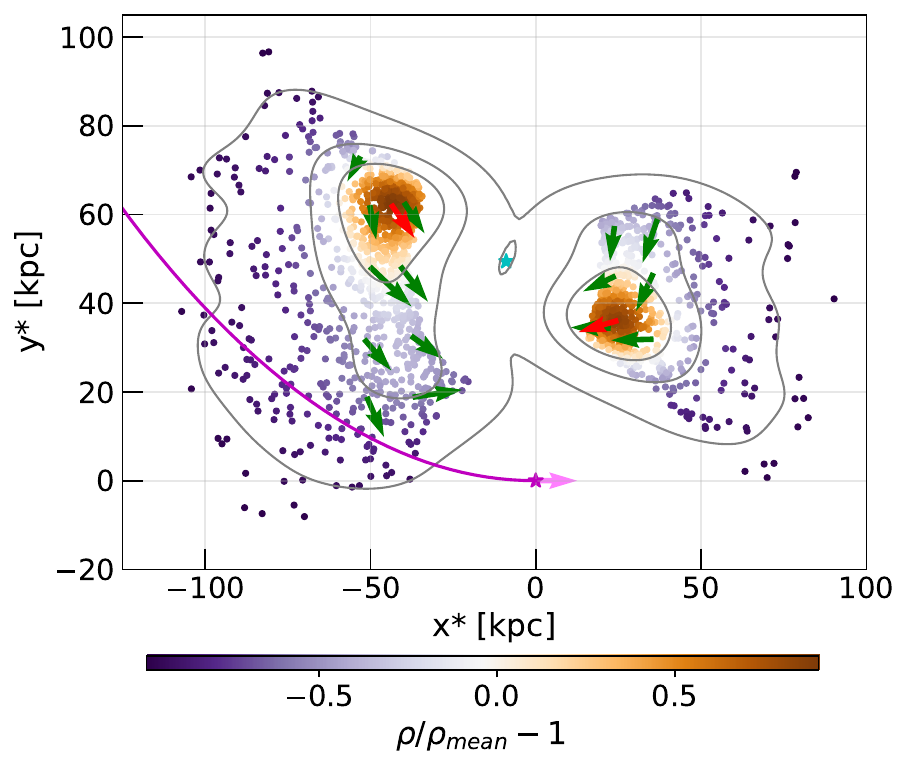}
    \includegraphics[width=\hsize]{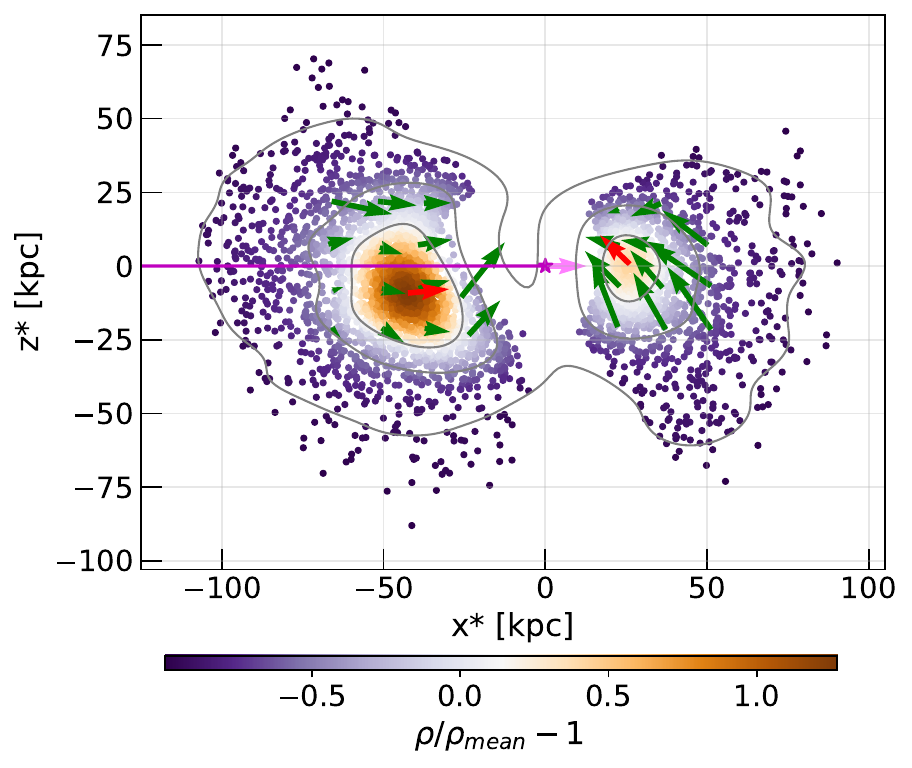}
    \caption{ Overdensity as a function of the position. Magenta line: past orbit of the CM. Magenta star: current position of the CM. Pink arrow: velocity of the CM. Red arrow: mean velocity of the wake or collective. Green arrow: mean velocity of a bin of 15 kpc. Cyan star: position of the galactic centre. Top panel: Orbit plane with $|z^*|<10 \,\rm kpc$. Bottom panel: $x^*-z^*$ plane (perpendicular to the orbit plane).}
    \label{wake_overdensity}
\end{figure}

 In Fig. \ref{wake_overdensity}, we plot the superficial overdensity of the wake along with the past orbit of the CM (magenta line), in a new coordinate frame; namely, the orbit frame (see appendix \ref{apendice:orbit}). In this frame, the plane $x^*-y^*$ contains the CM orbit and the $z^*$ axis is perpendicular to the CM orbit. The origin of this new coordinate frame is the current location of the CM, and the $x^*$ axis is coincident with the direction of the DM subhalo mass velocity. As one can see, the wake is located in $x^*<0$ and it moves towards the perturber. In particular, the lower panel of Fig. \ref{wake_overdensity} is in good agreement with the results presented in Fig. 1 of \citet{Buschmann:2018}.

The maximum value of the density is approximately  $x^*=-48.97$ kpc, $y^*=55.65$ kpc, and $z^*=-7.94$ kpc in the new orbit frame. Using the stars in a $10$ kpc neighbourhood (177 stars), we computed the velocity dispersion, resulting in $(48\pm 3)$ km/s. The characterisation of its complete stellar population and dynamical properties will be addressed in a forthcoming work.

The \textit{Gaia} satellite data and spectroscopic surveys have unveiled the structure, composition, and formation history of the MW's stellar halo in recent years \citep{Helmi:2018,Belokurov:2019, Kruijssen:2020,Callingham:2022}. Satellite galaxies, globular clusters, stars, and streams are now associated with different halo components, which constitute the remains of our galaxy's past merger events. 

\begin{figure}
    \centering
    \includegraphics[width=\hsize]{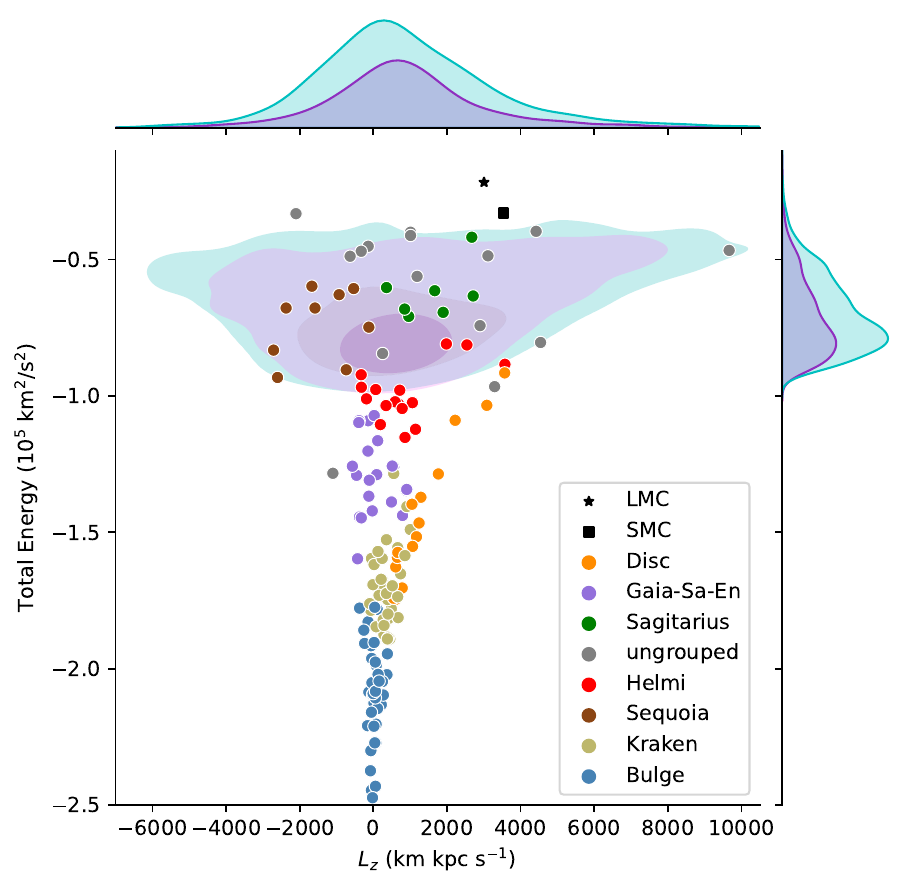}
    \caption{ $E$-$L_z$ diagram. The mergers data were taken from  \cite{Callingham:2022}. Cyan region: wake. Magenta region: collective response.}
    \label{E_LZ}
\end{figure}

In order to study the possible origin of the stellar populations of the wake and collective response, we have made an $E$ vs $L_z$ diagram (see Fig. \ref{E_LZ}). We have taken the same coordinate system convention as the one adopted by \citet{Callingham:2022}. In this diagram, we have discriminated the different known mergers of the MW (colour points, data extracted from \citet{Callingham:2022}), the LMC (star), the SMC (squared), and the wake and collective response (cyan and magenta regions, respectively). As one can see, both the wake and the collective are extended regions in the diagram without a defined sign of $L_z$ but limited in energy; however, it is not in the range of the Gaia-Sausage-Enceladus energy. The mean values for the wake are $E=-0.703 \times 10^5\, \rm{km}^2/\rm{s}^2$ and $L_z= 890.433 \,\rm{km\, kpc\, s}^{-1}$; meanwhile, for the collective, they are $E= -0.716\times 10^5\, \rm{km}^2/\rm{s}^2$ and $L_z= 902.391\, \rm{km\, kpc\, s}^{-1}$. 

It should be noted that the stellar population of the MW's Halo was accreted by mergers of older origin such as Sequoia, Saggitarius, Helmi, ED-3-4-5-6, Typhon, and L-RL64 \citep{Dodd:2023}. Consequently, the recent impact of the LMC could dynamically affect all these stellar populations; however, more studies are needed.

\begin{figure}[h!]
   \centering
   \includegraphics[width=\hsize]{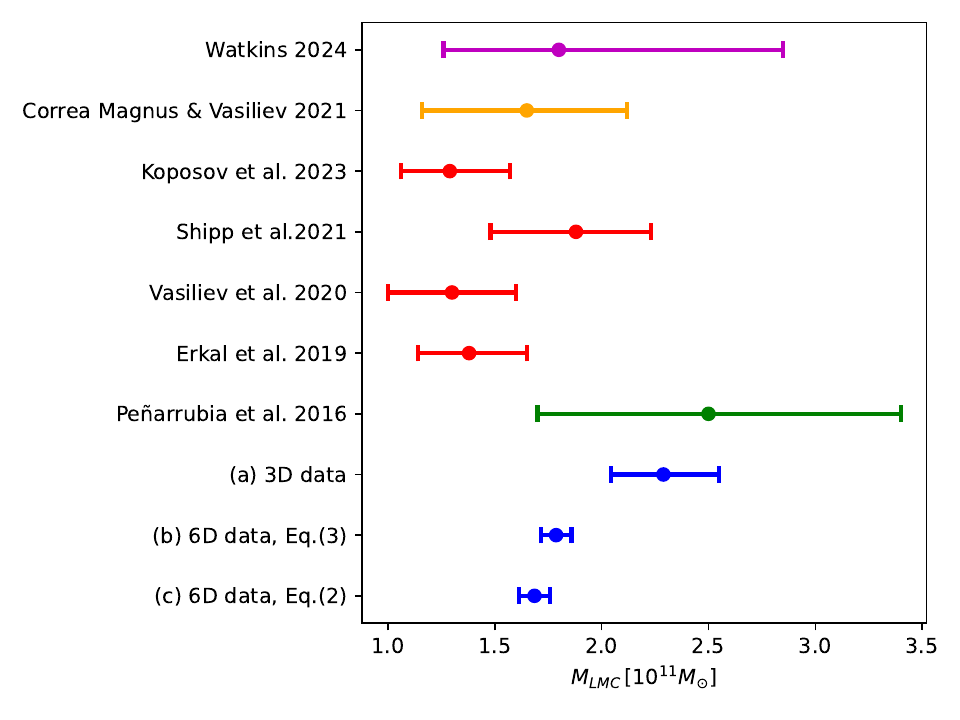}
      \caption{ Dark matter subhalo mass estimation of the LMC. Orange line: kinematic estimation from MW satellites. Red lines: estimation from stellar streams. Green line: estimation based on momentum balance in the Local Group. Blue lines: our results obtained from the likelihood analysis. Case (a): $3D$ data, $M_{LMC}=2.289^{+0.260}_{-0.240}\times 10^{11}\rm{M}_{\odot}$. Case (b): $6D$ data (Eq.\eqref{f1-3}), $M_{LMC}=1.787^{+0.072}_{-0.069}\times 10^{11}\rm{M}_{\odot}$. Case (c): $6D$ data (Eq.\eqref{f1-1}), $M_{LMC}=1.686^{+0.071}_{-0.072}\times 10^{11}\rm{M}_{\odot}$. The error bars of our results are purely statistical and based on a restricted one-parameter model. }
         \label{masas}
   \end{figure}

Next, we performed a statistical analysis in order to obtain the DM subhalo mass of the CM, and therefore the DM subhalo mass of the LMC. The radius for the region of interest was fixed at $R=100\, {\rm kpc}$, the subhalo velocity, $v_s$, was fixed at $314.23$ km/s \citep{vanderMarel:2002,Martinez:2019}, and $n_0$ was obtained from the reduced observational data described in the previous section. For the velocity dispersion, we performed a statistical analysis of the data and obtained the velocity standard deviation in each axis.
For the analysis using Eqs. \eqref{f1-1}, we considered the larger component of $\bar{v}_0$ in the calculations. We used the described density profile in Section \ref{like} and we present our results in Fig. \ref{masas}, along with the DM subhalo mass estimation of the LMC published in the literature \citep{Watkins:2024,koposov:2023,Shipp:2021,Vasiliev:2020,Erkal:2019,Penarrubia:2015} (using a different method, indicated with colours) along with our results (the last three values, with their corresponding statistical errors). 

 The DM subhalo mass was computed by using the space distribution function (case (a)) and the phase-space distribution function of  Eq.\eqref{f1-3} (case (b)) and Eq.\eqref{f1-1} (case (c)). Our results are consistent, despite the distribution function used in the analysis. However, the fit obtained using only the space data is slightly higher than the $6D$ analysis results. Furthermore, our findings agree within $3\sigma$ with the literature \citep{Vasiliev:2023}.
   
 We also performed the statistical analysis ($6D$ Eq.\eqref{f1-1}) using only the data with measured radial velocity \citep{GAIA:DR3,sloan:2012,wang:2022}. We found an LMC subhalo mass of $M_{LMC}=1.594^{+0.203}_{-0.196}\times 10^{11}\rm{M}_{\odot}$, which is in good agreement with our results.


\section{Conclusions}
\label{conclusiones}
In this work, we employed the recently published \textit{Gaia} Data Release 3, which improves the precision of proper motions along with the Segue catalogue \citep{sloan:2012} and the one provided by \citet{wang:2022}. This enabled us to extend the K-giant catalogue originally provided by \citet{Conroy:2021}, and also to construct a catalogue for RR-Lyrae stars, both in 6D data. We reproduced the previously published results and identified the overdensities associated with a wake and the collective response using these two halo tracers. A notable finding of this study is the extension of the southern overdensity towards lower galactocentric distances; that is, between 30 and 100 kpc. Moreover, we were able to show that the southern overdensity, identified as the wake, trails the CM of Magellanic Clouds (see Fig. \ref{wake_overdensity}).

We have confirmed that the Pisces plume overdensity, described in \cite{Belokurov2019-eb}, is associated with the wake of the Magellanic Clouds in the outer regions of the MW's halo. Furthermore, we have discovered that the overdensity on the halo's stellar population, caused by the Magellanic Clouds' wake and the global response, affects the stars in the MW's halo, regardless of which past merger event they were accreted from.

As per the theoretical proposal made by \citet{Buschmann:2018}, we were able to estimate the mass of the LMC DM subhalo for the first time by using \textit{Gaia} observational data. We found a reliable estimation of the DM halo surrounding the LMC by performing two different analyses, using only the space distribution data and using both the phase and space data. Considering a relationship between the Large and Small Magellanic Clouds' masses, our study has successfully determined the mass of the DM subhalo of the larger cloud. Even more, our findings are in agreement with prior results \citep{Correa:2021,koposov:2023,Shipp:2021,Vasiliev:2020,Erkal:2019,Penarrubia:2015}, within $3\sigma$. This consistency with previous studies indicates the reliability of our methodology. Additionally, this method gives competitive errors compared to different mass determination methods. It is important to point out that the errors mentioned were calculated using the assumptions of a Plummer profile for a spherical subhalo. However, if a more complex multi-parameter model such as an ellipsoidal Navarrro, Frenk $\&$ White (NFW) model were used, it is anticipated that the errors would increase.

\begin{acknowledgements}
 This work was partially supported by grants from the National Research Council of Argentina (CONICET PIP 616). K.J.F. is a Post Doctoral fellow of the CONICET. M.E.M. and M.D. are members of the Scientific Research Career of the CONICET. M.D. work was supported by the Preparing for Astrophysics with LSST Program, funded by the Heising Simons Foundation through grant 2021-2975, and administered by Las Cumbres Observatory.
 
We would like to express our gratitude to the referee for providing us with their valuable comments and suggestions.

M.D. thanks valuable suggestions by  Nicol\'as Garavito-Camargo and the support of the CCA and the Flatiron Institute.  M.D. also thanks the possibility of virtual participation in the following programs: "Building a physical understanding of galaxy evolution with data-driven astronomy" at the KITP-UCB, the Chicago "\textit{Gaia} DR3 Sprint" and  "Streams22: Community Atlas of Tidal Streams."  at the Carnegie Observatories and the Flatiron Institute. 

K.J.F wants to thank Alejo Molina Lera and Martín Gamboa Lerena for the useful discussions.
 
This work has made use of data from the European Space Agency (ESA) mission {\it Gaia}  (\url{https://www.cosmos.esa.int/gaia}), processed by the {\it Gaia}  Data Processing and Analysis Consortium (DPAC)
\url{https://www.cosmos.esa.int/web/gaia/dpac/consortium}), and code developed by the \textit{Gaia} Project Scientist Support Team. Funding for the DPAC has been provided by national institutions, in particular, the institutions participating in the {\it Gaia} Multilateral Agreement.

 This research has utilized the following software: Astropy \citep{astropy}, Matplotlib \citep{matplotlib}, Pandas \citep{pandas}, Seaborn \citep{seaborn}, Healpy \citep{healpy}, SciPy \citep{scipy}, NumPy \citep{numpy}, emcee \citep{emcee}, gala \citep{gala,gala2}, The Jupyter Notebook \citep{jupyter-notebook}, Scikit-learn \citep{scikit-learn}, Pzflow \citep{pzflow}, and TOPCAT \citep{topcat}.
\end{acknowledgements}

%
%
  \bibliographystyle{aa.bst} 
   \bibliography{bibliografia.bib}

\begin{thebibliography}{86}
\expandafter\ifx\csname natexlab\endcsname\relax\def\natexlab#1{#1}\fi

\bibitem[{{Abdalla} {et~al.}(2022){Abdalla}, {Aharonian}, {Benkhali},
  {Ang{\"u}ner}, {Armand}, {Ashkar}, {Backes}, {Baghmanyan}, {Martins},
  {Batzofin}, {Becherini}, {Berge}, {Bernl{\"o}hr}, {Bi}, {B{\"o}ttcher},
  {Bolmont}, {de Lavergne}, {Brose}, {Brun}, {Cangemi}, {Caroff}, {Cerruti},
  {Chand}, {Chen}, {Cotter}, {Mbarubucyeye}, {Devin}, {Djannati-Ata{\"\i}},
  {Dmytriiev}, {Doroshenko}, {Egberts}, {Fiasson}, {de Clairfontaine},
  {Fontaine}, {Funk}, {Gabici}, {Giavitto}, {Glawion}, {Glicenstein},
  {Grondin}, {Hinton}, {Hofmann}, {Holch}, {Holler}, {Horns}, {Huang},
  {Jamrozy}, {Jankowsky}, {Kasai}, {Katarzy{\'n}ski}, {Katz}, {Kh{\'e}lifi},
  {Klu{\'z}niak}, {Komin}, {Kosack}, {Kostunin}, {Lamanna}, {Lemoine-Goumard},
  {Lenain}, {Leuschner}, {Lohse}, {Luashvili}, {Lypova}, {Mackey}, {Malyshev},
  {Malyshev}, {Marandon}, {Marchegiani}, {Mart{\'\i}-Devesa}, {Marx}, {Maurin},
  {Meyer}, {Mitchell}, {Moderski}, {Montanari}, {Moulin}, {Muller}, {de
  Naurois}, {Niemiec}, {Noel}, {Ohm}, {Olivera-Nieto}, {Wilhelmi}, {Ostrowski},
  {Panny}, {Panter}, {Parsons}, {Peron}, {Poireau}, {Prokoph}, {P{\"u}hlhofer},
  {Punch}, {Quirrenbach}, {Reichherzer}, {Reimer}, {Reimer}, {Renaud},
  {Rieger}, {Rowell}, {Rudak}, {Ricarte}, {Ruiz-Velasco}, {Sahakian},
  {Salzmann}, {Santangelo}, {Sasaki}, {Sch{\"u}ssler}, {Schutte}, {Schwanke},
  {Senniappan}, {Shapopi}, {Sol}, {Specovius}, {Spencer}, {Stawarz},
  {Stegmann}, {Steinmassl}, {Steppa}, {Takahashi}, {Tanaka}, {Terrier},
  {Thorpe-Morgan}, {Tluczykont}, {Tsirou}, {Tsuji}, {Uchiyama}, {van Eldik},
  {Veh}, {Vink}, {Wagner}, {White}, {Wierzcholska}, {Wong}, {Zacharias},
  {Zargaryan}, {Zdziarski}, {Zech}, {Zhu}, {Zouari}, {{\.Z}ywucka}, \&
  {H.~E.~S.~S. Collaboration}}]{HEES:2022}
{Abdalla}, H., {Aharonian}, F., {Benkhali}, F.~A., {et~al.} 2022, \prl, 129,
  111101

\bibitem[{Abe {et~al.}(2023)Abe, Abe, Acciari, Aniello, Ansoldi, Antonelli,
  Arbet~Engels, Arcaro, Artero, Asano, Baack, Babi\ifmmode~\acute{c}\else
  \'{c}\fi{}, Baquero, Barres~de Almeida, Barrio,
  Batkovi\ifmmode~\acute{c}\else \'{c}\fi{}, Baxter, Becerra~Gonz\'alez,
  Bednarek, Bernardini, Bernardos, Berti, Besenrieder, Bhattacharyya,
  Bigongiari, Biland, Blanch, Bonnoli, Bo\ifmmode~\check{s}\else
  \v{s}\fi{}njak, Burelli, Busetto, Carosi, Carretero-Castrillo, Ceribella,
  Chai, Chilingarian, Cikota, Colombo, Contreras, Cortina, Covino, D'Amico,
  D'Elia, Da~Vela, Dazzi, De~Angelis, De~Lotto, Del~Popolo, Delfino, Delgado,
  Delgado~Mendez, Depaoli, Di~Pierro, Di~Venere, Do~Souto Espi\~neira,
  Dominis~Prester, Donini, Dorner, Doro, Elsaesser, Emery, Fallah~Ramazani,
  Fari\~na, Fattorini, Font, Fruck, Fukami, Fukazawa, Garc\'{\i}a~L\'opez,
  Garczarczyk, Gasparyan, Gaug, Giesbrecht~Paiva, Giglietto, Giordano, Gliwny,
  Godinovi\ifmmode~\acute{c}\else \'{c}\fi{}, Green, Green, Hadasch, Hahn,
  Hassan, Heckmann, Herrera, Hrupec, H\"utten, Imazawa, Inada, Iotov, Ishio,
  Jim\'enez~Mart\'{\i}nez, Jormanainen, Kerszberg, Kobayashi, Kubo, Kushida,
  Lamastra, Lelas, Leone, Lindfors, Linhoff, Lombardi, Longo, L\'opez-Coto,
  L\'opez-Moya, L\'opez-Oramas, Loporchio, Lorini, Lyard, Machado~de
  Oliveira~Fraga, Majumdar, Makariev, Maneva, Mang, Manganaro, Mangano,
  Mannheim, Mariotti, Mart\'{\i}nez, Mas~Aguilar, Mazin, Menchiari, Mender,
  Mi\ifmmode \acute{c}\else \'{c}\fi{}anovi\ifmmode~\acute{c}\else \'{c}\fi{},
  Miceli, Miener, Miranda, Mirzoyan, Molina, Mondal, Moralejo, Morcuende,
  Moreno, Nakamori, Nanci, Nava, Neustroev, Nievas~Rosillo, Nigro, Nilsson,
  Nishijima, Njoh~Ekoume, Noda, Nozaki, Ohtani, Oka, Otero-Santos, Paiano,
  Palatiello, Paneque, Paoletti, Paredes, Pavleti\ifmmode~\acute{c}\else
  \'{c}\fi{}, Persic, Pihet, Podobnik, Prada~Moroni, Prandini, Principe,
  Priyadarshi, Puljak, Rhode, Rib\'o, Rico, Righi, Rugliancich, Sahakyan,
  Saito, Sakurai, Satalecka, Saturni, Schleicher, Schmidt, Schmuckermaier,
  Schubert, Schweizer, Sitarek, Sliusar, Sobczynska, Spolon, Stamerra,
  Stri\ifmmode \check{s}\else \v{s}\fi{}kovi\ifmmode~\acute{c}\else \'{c}\fi{},
  Strom, Strzys, Suda, Suri\ifmmode~\acute{c}\else \'{c}\fi{}, Takahashi,
  Takeishi, Tavecchio, Temnikov, Terauchi, Terzi\ifmmode~\acute{c}\else
  \'{c}\fi{}, Teshima, Tosti, Truzzi, Tutone, Ubach, van Scherpenberg,
  Vazquez~Acosta, Ventura, Verguilov, Viale, Vigorito, Vitale, Vovk, Walter,
  Will, Wunderlich, Yamamoto, Zari\ifmmode~\acute{c}\else \'{c}\fi{},
  Hiroshima, \& Kohri}]{MAGIC:2023}
Abe, H., Abe, S., Acciari, V.~A., {et~al.} 2023, Phys. Rev. Lett., 130, 061002

\bibitem[{{Acharyya} {et~al.}(2023){Acharyya}, {Archer}, {Bangale},
  {Bartkoske}, {Batista}, {Baumgart}, {Benbow}, {Buckley}, {Falcone}, {Feng},
  {Finley}, {Foote}, {Fortson}, {Furniss}, {Gallagher}, {Hanlon}, {Hervet},
  {Hoang}, {Holder}, {Humensky}, {Jin}, {Kaaret}, {Kertzman}, {Kherlakian},
  {Kieda}, {Kleiner}, {Korzoun}, {Krennrich}, {Lang}, {Lundy}, {Maier},
  {McGrath}, {Moriarty}, {O'Brien}, {Ong}, {Pfrang}, {Pohl}, {Pueschel},
  {Quinn}, {Ragan}, {Reynolds}, {Roache}, {Rodd}, {Ryan}, {Sadeh}, {Saha},
  {Santander}, {Sembroski}, {Shang}, {Splettstoesser}, {Tak}, {Tucci},
  {Vassiliev}, \& {Williams}}]{VERITAS:2023}
{Acharyya}, A., {Archer}, A., {Bangale}, P., {et~al.} 2023, \apj, 945, 101

\bibitem[{{Aguilar-Arg{\"u}ello} {et~al.}(2022){Aguilar-Arg{\"u}ello},
  {Valenzuela}, \& {Trelles}}]{Aguilar:2022}
{Aguilar-Arg{\"u}ello}, G., {Valenzuela}, O., \& {Trelles}, A. 2022, \aap, 663,
  A93

\bibitem[{{Ahn} {et~al.}(2012){Ahn}, {Alexandroff}, {Allende Prieto},
  {Anderson}, {Anderton}, {Andrews}, {Aubourg}, {Bailey}, {Balbinot}, {Barnes},
  {Bautista}, {Beers}, {Beifiori}, {Berlind}, {Bhardwaj}, {Bizyaev}, {Blake},
  {Blanton}, {Blomqvist}, {Bochanski}, {Bolton}, {Borde}, {Bovy}, {Brandt},
  {Brinkmann}, {Brown}, {Brownstein}, {Bundy}, {Busca}, {Carithers}, {Carnero},
  {Carr}, {Casetti-Dinescu}, {Chen}, {Chiappini}, {Comparat}, {Connolly},
  {Crepp}, {Cristiani}, {Croft}, {Cuesta}, {da Costa}, {Davenport}, {Dawson},
  {de Putter}, {De Lee}, {Delubac}, {Dhital}, {Ealet}, {Ebelke}, {Edmondson},
  {Eisenstein}, {Escoffier}, {Esposito}, {Evans}, {Fan}, {Femen{\'\i}a
  Castell{\'a}}, {Fern{\'a}ndez Alvar}, {Ferreira}, {Filiz Ak}, {Finley},
  {Fleming}, {Font-Ribera}, {Frinchaboy}, {Garc{\'\i}a-Hern{\'a}ndez},
  {Garc{\'\i}a P{\'e}rez}, {Ge}, {G{\'e}nova-Santos}, {Gillespie}, {Girardi},
  {Gonz{\'a}lez Hern{\'a}ndez}, {Grebel}, {Gunn}, {Guo}, {Haggard}, {Hamilton},
  {Harris}, {Hawley}, {Hearty}, {Ho}, {Hogg}, {Holtzman}, {Honscheid},
  {Huehnerhoff}, {Ivans}, {Ivezi{\'c}}, {Jacobson}, {Jiang}, {Johansson},
  {Johnson}, {Kauffmann}, {Kirkby}, {Kirkpatrick}, {Klaene}, {Knapp}, {Kneib},
  {Le Goff}, {Leauthaud}, {Lee}, {Lee}, {Long}, {Loomis}, {Lucatello},
  {Lundgren}, {Lupton}, {Ma}, {Ma}, {MacDonald}, {Mack}, {Mahadevan}, {Maia},
  {Majewski}, {Makler}, {Malanushenko}, {Malanushenko}, {Manchado},
  {Mandelbaum}, {Manera}, {Maraston}, {Margala}, {Martell}, {McBride},
  {McGreer}, {McMahon}, {M{\'e}nard}, {Meszaros}, {Miralda-Escud{\'e}},
  {Montero-Dorta}, {Montesano}, {Morrison}, {Muna}, {Munn}, {Murayama},
  {Myers}, {Neto}, {Nguyen}, {Nichol}, {Nidever}, {Noterdaeme}, {Nuza},
  {Ogando}, {Olmstead}, {Oravetz}, {Owen}, {Padmanabhan},
  {Palanque-Delabrouille}, {Pan}, {Parejko}, {Parihar}, {P{\^a}ris},
  {Pattarakijwanich}, {Pepper}, {Percival}, {P{\'e}rez-Fournon},
  {P{\'e}rez-R{\`a}fols}, {Petitjean}, {Pforr}, {Pieri}, {Pinsonneault}, {Porto
  de Mello}, {Prada}, {Price-Whelan}, {Raddick}, {Rebolo}, {Rich}, {Richards},
  {Robin}, {Rocha-Pinto}, {Rockosi}, {Roe}, {Ross}, {Ross}, {Rossi},
  {Rubi{\~n}o-Martin}, {Samushia}, {Sanchez Almeida}, {S{\'a}nchez},
  {Santiago}, {Sayres}, {Schlegel}, {Schlesinger}, {Schmidt}, {Schneider},
  {Schultheis}, {Schwope}, {Sc{\'o}ccola}, {Seljak}, {Sheldon}, {Shen}, {Shu},
  {Simmerer}, {Simmons}, {Skibba}, {Skrutskie}, {Slosar}, {Sobreira}, {Sobeck},
  {Stassun}, {Steele}, {Steinmetz}, {Strauss}, {Streblyanska}, {Suzuki},
  {Swanson}, {Tal}, {Thakar}, {Thomas}, {Thompson}, {Tinker}, {Tojeiro},
  {Tremonti}, {Vargas Maga{\~n}a}, {Verde}, {Viel}, {Vikas}, {Vogt}, {Wake},
  {Wang}, {Weaver}, {Weinberg}, {Weiner}, {West}, {White}, {Wilson},
  {Wisniewski}, {Wood-Vasey}, {Yanny}, {Y{\`e}che}, {York}, {Zamora},
  {Zasowski}, {Zehavi}, {Zhao}, {Zheng}, {Zhu}, \& {Zinn}}]{sloan:2012}
{Ahn}, C.~P., {Alexandroff}, R., {Allende Prieto}, C., {et~al.} 2012, \apjs,
  203, 21

\bibitem[{{Amar{\'e}} {et~al.}(2022){Amar{\'e}}, {Cebri{\'a}n}, {Cintas},
  {Coarasa}, {Garc{\'\i}a}, {Ortiz de Sol{\'o}rzano}, {Puimed{\'o}n},
  {Salinas}, {Sarsa}, {Villar}, {Mart{\'\i}nez}, {Oliv{\'a}n}, \&
  {Ortigoza}}]{ANAIS:2022}
{Amar{\'e}}, J., {Cebri{\'a}n}, S., {Cintas}, D., {et~al.} 2022, Moscow
  University Physics Bulletin, 77, 322

\bibitem[{{Aprile} {et~al.}(2023){Aprile}, {Abe}, {Agostini}, {Ahmed Maouloud},
  {Althueser}, {Andrieu}, {Angelino}, {Angevaare}, {Antochi}, {Ant{\'o}n
  Martin}, {Arneodo}, {Baudis}, {Baxter}, {Bazyk}, {Bellagamba}, {Biondi},
  {Bismark}, {Brookes}, {Brown}, {Bruenner}, {Bruno}, {Budnik}, {Bui}, {Cai},
  {Cardoso}, {Cichon}, {Cimental Chavez}, {Colijn}, {Conrad},
  {Cuenca-Garc{\'\i}a}, {Cussonneau}, {D'Andrea}, {Decowski}, {di Gangi}, {di
  Pede}, {Diglio}, {Eitel}, {Elykov}, {Farrell}, {Ferella}, {Ferrari},
  {Fischer}, {Flierman}, {Fulgione}, {Fuselli}, {Gaemers}, {Gaior}, {Gallo
  Rosso}, {Galloway}, {Gao}, {Glade-Beucke}, {Grandi}, {Grigat}, {Guan},
  {Guida}, {Hammann}, {Higuera}, {Hils}, {Hoetzsch}, {Hood}, {Howlett},
  {Iacovacci}, {Itow}, {Jakob}, {Joerg}, {Joy}, {Kato}, {Kara}, {Kavrigin},
  {Kazama}, {Kobayashi}, {Koltman}, {Kopec}, {Kuger}, {Landsman}, {Lang},
  {Levinson}, {Li}, {Li}, {Liang}, {Lindemann}, {Lindner}, {Liu}, {Loizeau},
  {Lombardi}, {Long}, {Lopes}, {Ma}, {Macolino}, {Mahlstedt}, {Mancuso},
  {Manenti}, {Marignetti}, {Marrod{\'a}n Undagoitia}, {Martens}, {Masbou},
  {Masson}, {Masson}, {Mastroianni}, {Messina}, {Miuchi}, {Mizukoshi},
  {Molinario}, {Moriyama}, {Mor{\^a}}, {Mosbacher}, {Murra}, {M{\"u}ller},
  {Ni}, {Oberlack}, {Paetsch}, {Palacio}, {Peres}, {Peters}, {Pienaar},
  {Pierre}, {Pizzella}, {Plante}, {Qi}, {Qin}, {Ram{\'\i}rez Garc{\'\i}a},
  {Singh}, {Sanchez}, {Dos Santos}, {Sarnoff}, {Sartorelli}, {Schreiner},
  {Schulte}, {Schulte}, {Schulze Ei{\ss}ing}, {Schumann}, {Scotto Lavina},
  {Selvi}, {Semeria}, {Shagin}, {Shi}, {Shockley}, {Silva}, {Simgen}, {Takeda},
  {Tan}, {Terliuk}, {Thers}, {Toschi}, {Trinchero}, {Tunnell}, {T{\"o}nnies},
  {Valerius}, {Volta}, {Weinheimer}, {Weiss}, {Wenz}, {Wittweg}, {Wolf}, {Wu},
  {Xing}, {Xu}, {Xu}, {Yamashita}, {Yang}, {Ye}, {Yuan}, {Zavattini}, {Zhong},
  {Zhu}, \& {Xenon Collaboration}}]{Aprile:2023}
{Aprile}, E., {Abe}, K., {Agostini}, F., {et~al.} 2023, \prl, 131, 041003

\bibitem[{{Astropy Collaboration} {et~al.}(2013){Astropy Collaboration},
  {Robitaille}, {Tollerud}, {Greenfield}, {Droettboom}, {Bray}, {Aldcroft},
  {Davis}, {Ginsburg}, {Price-Whelan}, {Kerzendorf}, {Conley}, {Crighton},
  {Barbary}, {Muna}, {Ferguson}, {Grollier}, {Parikh}, {Nair}, {Unther},
  {Deil}, {Woillez}, {Conseil}, {Kramer}, {Turner}, {Singer}, {Fox}, {Weaver},
  {Zabalza}, {Edwards}, {Azalee Bostroem}, {Burke}, {Casey}, {Crawford},
  {Dencheva}, {Ely}, {Jenness}, {Labrie}, {Lim}, {Pierfederici}, {Pontzen},
  {Ptak}, {Refsdal}, {Servillat}, \& {Streicher}}]{astropy}
{Astropy Collaboration}, {Robitaille}, T.~P., {Tollerud}, E.~J., {et~al.} 2013,
  \aap, 558, A33

\bibitem[{{Barberio} {et~al.}(2023){Barberio}, {Baroncelli}, {Bignell},
  {Bolognino}, {Brooks}, {Dastgiri}, {D'Imperio}, {Di Giacinto}, {Duffy},
  {Froehlich}, {Fu}, {Gerathy}, {Hill}, {Krishnan}, {Lane}, {Lawrence},
  {Leaver}, {Mahmood}, {Mariani}, {McGee}, {McKie}, {McNamara}, {Mews},
  {Melbourne}, {Milana}, {Milligan}, {Mould}, {Nuti}, {Pettinacci}, {Scutti},
  {Slavkovsk{\'a}}, {Spinks}, {Stanley}, {Stuchbery}, {Taylor}, {Tomei},
  {Urquijo}, {Vignoli}, {Williams}, {Zhong}, \& {Zurowski}}]{Barberio:2022}
{Barberio}, E., {Baroncelli}, T., {Bignell}, L.~J., {et~al.} 2023, European
  Physical Journal C, 83, 878

\bibitem[{{Bazarov} {et~al.}(2022){Bazarov}, {Benito}, {H{\"u}tsi}, {Kipper},
  {Pata}, \& {P{\~o}der}}]{Bazarov:2022}
{Bazarov}, A., {Benito}, M., {H{\"u}tsi}, G., {et~al.} 2022, Astronomy and
  Computing, 41, 100667

\bibitem[{Belokurov {et~al.}(2019)Belokurov, Deason, Erkal, Koposov,
  Carballo-Bello, Smith, Jethwa, \& Navarrete}]{Belokurov:2019}
Belokurov, V., Deason, A.~J., Erkal, D., {et~al.} 2019, Monthly Notices of the
  Royal Astronomical Society: Letters, 488, L47

\bibitem[{{Belokurov} {et~al.}(2019){Belokurov}, {Deason}, {Erkal}, {Koposov},
  {Carballo-Bello}, {Smith}, {Jethwa}, \& {Navarrete}}]{Belokurov2019-eb}
{Belokurov}, V., {Deason}, A.~J., {Erkal}, D., {et~al.} 2019, \mnras, 488, L47

\bibitem[{{Bennett} \& {Bovy}(2019)}]{Bennett}
{Bennett}, M. \& {Bovy}, J. 2019, \mnras, 482, 1417

\bibitem[{{Bernabei} {et~al.}(2022){Bernabei}, {Belli}, {Bussolotti},
  {Cerulli}, {di Marco}, {Merlo}, {Montecchia}, {Cappella}, {D'Angelo},
  {Incicchitti}, {Mattei}, {Caracciolo}, {Dai}, {He}, {Ma}, {Sheng}, \&
  {Ye}}]{Bernabei:2022}
{Bernabei}, R., {Belli}, P., {Bussolotti}, A., {et~al.} 2022, in The Fifteenth
  Marcel Grossmann Meeting on General Relativity. Edited by E. S. Battistelli,
  ed. E.~S. {Battistelli}, R.~T. {Jantzen}, \& R.~{Ruffini}, 1285--1290

\bibitem[{{Bovy}(2015)}]{Bovy:2015}
{Bovy}, J. 2015, \apjs, 216, 29

\bibitem[{{Breiman}(2001)}]{Breiman:2001}
{Breiman}, L. 2001, Machine Learning, 45, 5

\bibitem[{Buehler \& Desjacques(2023)}]{Buehler:2023}
Buehler, R. \& Desjacques, V. 2023, Phys. Rev. D, 107, 023516

\bibitem[{{Buschmann} {et~al.}(2018){Buschmann}, {Kopp}, {Safdi}, \&
  {Wu}}]{Buschmann:2018}
{Buschmann}, M., {Kopp}, J., {Safdi}, B.~R., \& {Wu}, C.-L. 2018, \prl, 120,
  211101

\bibitem[{{Callingham} {et~al.}(2022){Callingham}, {Cautun}, {Deason}, {Frenk},
  {Grand}, \& {Marinacci}}]{Callingham:2022}
{Callingham}, T.~M., {Cautun}, M., {Deason}, A.~J., {et~al.} 2022, \mnras, 513,
  4107

\bibitem[{Carr \& K\"{u}hnel(2020)}]{doi:10.1146/annurev-nucl-050520-125911}
Carr, B. \& K\"{u}hnel, F. 2020, Annual Review of Nuclear and Particle Science,
  70, 355

\bibitem[{Chandra {et~al.}(2023{\natexlab{a}})Chandra, Naidu, Conroy, Bonaca,
  Zaritsky, Cargile, Caldwell, Johnson, Han, \& Ting}]{Chandra:2023a}
Chandra, V., Naidu, R.~P., Conroy, C., {et~al.} 2023{\natexlab{a}}, The
  Astrophysical Journal, 956, 110

\bibitem[{Chandra {et~al.}(2023{\natexlab{b}})Chandra, Naidu, Conroy, Ji, Rix,
  Bonaca, Cargile, Han, Johnson, Ting, Woody, \& Zaritsky}]{Chandra:2023b}
Chandra, V., Naidu, R.~P., Conroy, C., {et~al.} 2023{\natexlab{b}}, The
  Astrophysical Journal, 951, 26

\bibitem[{{Choi} {et~al.}(2016){Choi}, {Dotter}, {Conroy}, {Cantiello},
  {Paxton}, \& {Johnson}}]{MESA_2}
{Choi}, J., {Dotter}, A., {Conroy}, C., {et~al.} 2016, \apj, 823, 102

\bibitem[{Clowe {et~al.}(2006)}]{Clowe:2006}
Clowe, D. {et~al.} 2006, Astrophys. J., 648, L109

\bibitem[{{Conroy} {et~al.}(2021){Conroy}, {Naidu}, {Garavito-Camargo},
  {Besla}, {Zaritsky}, {Bonaca}, \& {Johnson}}]{Conroy:2021}
{Conroy}, C., {Naidu}, R.~P., {Garavito-Camargo}, N., {et~al.} 2021, \nat, 592,
  534

\bibitem[{Correa Magnus \& Vasiliev(2021)}]{Correa:2021}
Correa Magnus, L. \& Vasiliev, E. 2021, Monthly Notices of the Royal
  Astronomical Society, 511, 2610

\bibitem[{{Craig} {et~al.}(2022){Craig}, {Chakrabarti}, {Baum}, \&
  {Lewis}}]{Craig:2021}
{Craig}, P.~A., {Chakrabarti}, S., {Baum}, S., \& {Lewis}, B.~T. 2022, \mnras,
  517, 1737

\bibitem[{{Crenshaw} {et~al.}(2021){Crenshaw}, {Connolly}, \&
  {Kalmbach}}]{2021AAS...23823001C}
{Crenshaw}, J.~F., {Connolly}, A., \& {Kalmbach}, B. 2021, in American
  Astronomical Society Meeting Abstracts, Vol.~53, American Astronomical
  Society Meeting Abstracts, 230.01

\bibitem[{Crenshaw {et~al.}(2024)}]{pzflow}
Crenshaw, J.~F. {et~al.} 2024, jfcrenshaw/pzflow: v3.1.3

\bibitem[{Cunningham {et~al.}(2020)Cunningham, Garavito-Camargo, Deason,
  Johnston, Erkal, Laporte, Besla, Luger, \& Sanderson}]{Cunningham:2020}
Cunningham, E.~C., Garavito-Camargo, N., Deason, A.~J., {et~al.} 2020, The
  Astrophysical Journal, 898, 4

\bibitem[{{Dodd} {et~al.}(2023){Dodd}, {Callingham}, {Helmi}, {Matsuno},
  {Ruiz-Lara}, {Balbinot}, \& {L{\"o}vdal}}]{Dodd:2023}
{Dodd}, E., {Callingham}, T.~M., {Helmi}, A., {et~al.} 2023, \aap, 670, L2

\bibitem[{{Dotter}(2016)}]{MESA_1}
{Dotter}, A. 2016, \apjs, 222, 8

\bibitem[{{Drimmel} \& {Poggio}(2018)}]{Drimmel}
{Drimmel}, R. \& {Poggio}, E. 2018, Research Notes of the American Astronomical
  Society, 2, 210

\bibitem[{{Drlica-Wagner} {et~al.}(2020{\natexlab{a}}){Drlica-Wagner},
  {Bechtol}, {Mau}, {McNanna}, {Nadler}, {Pace}, {Li}, {Pieres}, {Rozo},
  {Simon}, {Walker}, {Wechsler}, {Abbott}, {Allam}, {Annis}, {Bertin},
  {Brooks}, {Burke}, {Rosell}, {Carrasco Kind}, {Carretero}, {Costanzi}, {da
  Costa}, {De Vicente}, {Desai}, {Diehl}, {Doel}, {Eifler}, {Everett},
  {Flaugher}, {Frieman}, {Garc{\'\i}a-Bellido}, {Gaztanaga}, {Gruen},
  {Gruendl}, {Gschwend}, {Gutierrez}, {Honscheid}, {James}, {Krause}, {Kuehn},
  {Kuropatkin}, {Lahav}, {Maia}, {Marshall}, {Melchior}, {Menanteau}, {Miquel},
  {Palmese}, {Plazas}, {Sanchez}, {Scarpine}, {Schubnell}, {Serrano},
  {Sevilla-Noarbe}, {Smith}, {Suchyta}, {Tarle}, \& {DES
  Collaboration}}]{Drlica-Wagner:2020}
{Drlica-Wagner}, A., {Bechtol}, K., {Mau}, S., {et~al.} 2020{\natexlab{a}},
  \apj, 893, 47

\bibitem[{{Drlica-Wagner} {et~al.}(2020{\natexlab{b}}){Drlica-Wagner},
  {Bechtol}, {Mau}, {McNanna}, {Nadler}, {Pace}, {Li}, {Pieres}, {Rozo},
  {Simon}, {Walker}, {Wechsler}, {Abbott}, {Allam}, {Annis}, {Bertin},
  {Brooks}, {Burke}, {Rosell}, {Carrasco Kind}, {Carretero}, {Costanzi}, {da
  Costa}, {De Vicente}, {Desai}, {Diehl}, {Doel}, {Eifler}, {Everett},
  {Flaugher}, {Frieman}, {Garc{\'\i}a-Bellido}, {Gaztanaga}, {Gruen},
  {Gruendl}, {Gschwend}, {Gutierrez}, {Honscheid}, {James}, {Krause}, {Kuehn},
  {Kuropatkin}, {Lahav}, {Maia}, {Marshall}, {Melchior}, {Menanteau}, {Miquel},
  {Palmese}, {Plazas}, {Sanchez}, {Scarpine}, {Schubnell}, {Serrano},
  {Sevilla-Noarbe}, {Smith}, {Suchyta}, {Tarle}, \& {DES
  Collaboration}}]{galaxias}
{Drlica-Wagner}, A., {Bechtol}, K., {Mau}, S., {et~al.} 2020{\natexlab{b}},
  \apj, 893, 47

\bibitem[{Durkan {et~al.}(2019)Durkan, Bekasov, Murray, \&
  Papamakarios}]{durkan2019neural}
Durkan, C., Bekasov, A., Murray, I., \& Papamakarios, G. 2019, Neural Spline
  Flows

\bibitem[{{Erkal} {et~al.}(2019){Erkal}, {Belokurov}, {Laporte}, {Koposov},
  {Li}, {Grillmair}, {Kallivayalil}, {Price-Whelan}, {Evans}, {Hawkins},
  {Hendel}, {Mateu}, {Navarro}, {del Pino}, {Slater}, {Sohn}, \& {Orphan Aspen
  Treasury Collaboration}}]{Erkal:2019}
{Erkal}, D., {Belokurov}, V., {Laporte}, C.~F.~P., {et~al.} 2019, \mnras, 487,
  2685

\bibitem[{{Foote} {et~al.}(2023){Foote}, {Besla}, {Mocz}, {Garavito-Camargo},
  {Lancaster}, {Sparre}, {Cunningham}, {Vogelsberger}, {G{\'o}mez}, \&
  {Laporte}}]{Foote:2023}
{Foote}, H.~R., {Besla}, G., {Mocz}, P., {et~al.} 2023, \apj, 954, 163

\bibitem[{{Foreman-Mackey} {et~al.}(2013){Foreman-Mackey}, {Hogg}, {Lang}, \&
  {Goodman}}]{emcee}
{Foreman-Mackey}, D., {Hogg}, D.~W., {Lang}, D., \& {Goodman}, J. 2013, \pasp,
  125, 306

\bibitem[{Furlanetto \& Loeb(2002)}]{Furlanetto:2002}
Furlanetto, S.~R. \& Loeb, A. 2002, The Astrophysical Journal, 565, 854

\bibitem[{{Gaia Collaboration} {et~al.}(2016){Gaia Collaboration}, {Prusti},
  {de Bruijne}, {Brown}, {Vallenari}, {Babusiaux}, {Bailer-Jones}, {Bastian},
  {Biermann}, {Evans}, {Eyer}, {Jansen}, {Jordi}, {Klioner}, {Lammers},
  {Lindegren}, {Luri}, {Mignard}, {Milligan}, {Panem}, {Poinsignon},
  {Pourbaix}, {Randich}, {Sarri}, {Sartoretti}, {Siddiqui}, {Soubiran},
  {Valette}, {van Leeuwen}, {Walton}, {Aerts}, {Arenou}, {Cropper}, {Drimmel},
  {H{\o}g}, {Katz}, {Lattanzi}, {O'Mullane}, {Grebel}, {Holland}, {Huc},
  {Passot}, {Bramante}, {Cacciari}, {Casta{\~n}eda}, {Chaoul}, {Cheek}, {De
  Angeli}, {Fabricius}, {Guerra}, {Hern{\'a}ndez}, {Jean-Antoine-Piccolo},
  {Masana}, {Messineo}, {Mowlavi}, {Nienartowicz}, {Ord{\'o}{\~n}ez-Blanco},
  {Panuzzo}, {Portell}, {Richards}, {Riello}, {Seabroke}, {Tanga},
  {Th{\'e}venin}, {Torra}, {Els}, {Gracia-Abril}, {Comoretto},
  {Garcia-Reinaldos}, {Lock}, {Mercier}, {Altmann}, {Andrae}, {Astraatmadja},
  {Bellas-Velidis}, {Benson}, {Berthier}, {Blomme}, {Busso}, {Carry},
  {Cellino}, {Clementini}, {Cowell}, {Creevey}, {Cuypers}, {Davidson}, {De
  Ridder}, {de Torres}, {Delchambre}, {Dell'Oro}, {Ducourant}, {Fr{\'e}mat},
  {Garc{\'\i}a-Torres}, {Gosset}, {Halbwachs}, {Hambly}, {Harrison}, {Hauser},
  {Hestroffer}, {Hodgkin}, {Huckle}, {Hutton}, {Jasniewicz}, {Jordan},
  {Kontizas}, {Korn}, {Lanzafame}, {Manteiga}, {Moitinho}, {Muinonen},
  {Osinde}, {Pancino}, {Pauwels}, {Petit}, {Recio-Blanco}, {Robin}, {Sarro},
  {Siopis}, {Smith}, {Smith}, {Sozzetti}, {Thuillot}, {van Reeven}, {Viala},
  {Abbas}, {Abreu Aramburu}, {Accart}, {Aguado}, {Allan}, {Allasia},
  {Altavilla}, {{\'A}lvarez}, {Alves}, {Anderson}, {Andrei}, {Anglada Varela},
  {Antiche}, {Antoja}, {Ant{\'o}n}, {Arcay}, {Atzei}, {Ayache}, {Bach},
  {Baker}, {Balaguer-N{\'u}{\~n}ez}, {Barache}, {Barata}, {Barbier}, {Barblan},
  {Baroni}, {Barrado y Navascu{\'e}s}, {Barros}, {Barstow}, {Becciani},
  {Bellazzini}, {Bellei}, {Bello Garc{\'\i}a}, {Belokurov}, {Bendjoya},
  {Berihuete}, {Bianchi}, {Bienaym{\'e}}, {Billebaud}, {Blagorodnova},
  {Blanco-Cuaresma}, {Boch}, {Bombrun}, {Borrachero}, {Bouquillon}, {Bourda},
  {Bouy}, {Bragaglia}, {Breddels}, {Brouillet}, {Br{\"u}semeister},
  {Bucciarelli}, {Budnik}, {Burgess}, {Burgon}, {Burlacu}, {Busonero}, {Buzzi},
  {Caffau}, {Cambras}, {Campbell}, {Cancelliere}, {Cantat-Gaudin}, {Carlucci},
  {Carrasco}, {Castellani}, {Charlot}, {Charnas}, {Charvet}, {Chassat},
  {Chiavassa}, {Clotet}, {Cocozza}, {Collins}, {Collins}, {Costigan}, {Crifo},
  {Cross}, {Crosta}, {Crowley}, {Dafonte}, {Damerdji}, {Dapergolas}, {David},
  {David}, {De Cat}, {de Felice}, {de Laverny}, {De Luise}, {De March}, {de
  Martino}, {de Souza}, {Debosscher}, {del Pozo}, {Delbo}, {Delgado},
  {Delgado}, {di Marco}, {Di Matteo}, {Diakite}, {Distefano}, {Dolding}, {Dos
  Anjos}, {Drazinos}, {Dur{\'a}n}, {Dzigan}, {Ecale}, {Edvardsson}, {Enke},
  {Erdmann}, {Escolar}, {Espina}, {Evans}, {Eynard Bontemps}, {Fabre},
  {Fabrizio}, {Faigler}, {Falc{\~a}o}, {Farr{\`a}s Casas}, {Faye}, {Federici},
  {Fedorets}, {Fern{\'a}ndez-Hern{\'a}ndez}, {Fernique}, {Fienga}, {Figueras},
  {Filippi}, {Findeisen}, {Fonti}, {Fouesneau}, {Fraile}, {Fraser}, {Fuchs},
  {Furnell}, {Gai}, {Galleti}, {Galluccio}, {Garabato}, {Garc{\'\i}a-Sedano},
  {Gar{\'e}}, {Garofalo}, {Garralda}, {Gavras}, {Gerssen}, {Geyer}, {Gilmore},
  {Girona}, {Giuffrida}, {Gomes}, {Gonz{\'a}lez-Marcos},
  {Gonz{\'a}lez-N{\'u}{\~n}ez}, {Gonz{\'a}lez-Vidal}, {Granvik}, {Guerrier},
  {Guillout}, {Guiraud}, {G{\'u}rpide}, {Guti{\'e}rrez-S{\'a}nchez}, {Guy},
  {Haigron}, {Hatzidimitriou}, {Haywood}, {Heiter}, {Helmi}, {Hobbs},
  {Hofmann}, {Holl}, {Holland}, {Hunt}, {Hypki}, {Icardi}, {Irwin}, {Jevardat
  de Fombelle}, {Jofr{\'e}}, {Jonker}, {Jorissen}, {Julbe}, {Karampelas},
  {Kochoska}, {Kohley}, {Kolenberg}, {Kontizas}, {Koposov}, {Kordopatis},
  {Koubsky}, {Kowalczyk}, {Krone-Martins}, {Kudryashova}, {Kull}, {Bachchan},
  {Lacoste-Seris}, {Lanza}, {Lavigne}, {Le Poncin-Lafitte}, {Lebreton},
  {Lebzelter}, {Leccia}, {Leclerc}, {Lecoeur-Taibi}, {Lemaitre}, {Lenhardt},
  {Leroux}, {Liao}, {Licata}, {Lindstr{\o}m}, {Lister}, {Livanou}, {Lobel},
  {L{\"o}ffler}, {L{\'o}pez}, {Lopez-Lozano}, {Lorenz}, {Loureiro},
  {MacDonald}, {Magalh{\~a}es Fernandes}, {Managau}, {Mann}, {Mantelet},
  {Marchal}, {Marchant}, {Marconi}, {Marie}, {Marinoni}, {Marrese},
  {Marschalk{\'o}}, {Marshall}, {Mart{\'\i}n-Fleitas}, {Martino}, {Mary},
  {Matijevi{\v{c}}}, {Mazeh}, {McMillan}, {Messina}, {Mestre}, {Michalik},
  {Millar}, {Miranda}, {Molina}, {Molinaro}, {Molinaro}, {Moln{\'a}r},
  {Moniez}, {Montegriffo}, {Monteiro}, {Mor}, {Mora}, {Morbidelli}, {Morel},
  {Morgenthaler}, {Morley}, {Morris}, {Mulone}, {Muraveva}, {Musella},
  {Narbonne}, {Nelemans}, {Nicastro}, {Noval}, {Ord{\'e}novic},
  {Ordieres-Mer{\'e}}, {Osborne}, {Pagani}, {Pagano}, {Pailler}, {Palacin},
  {Palaversa}, {Parsons}, {Paulsen}, {Pecoraro}, {Pedrosa}, {Pentik{\"a}inen},
  {Pereira}, {Pichon}, {Piersimoni}, {Pineau}, {Plachy}, {Plum}, {Poujoulet},
  {Pr{\v{s}}a}, {Pulone}, {Ragaini}, {Rago}, {Rambaux}, {Ramos-Lerate},
  {Ranalli}, {Rauw}, {Read}, {Regibo}, {Renk}, {Reyl{\'e}}, {Ribeiro},
  {Rimoldini}, {Ripepi}, {Riva}, {Rixon}, {Roelens}, {Romero-G{\'o}mez},
  {Rowell}, {Royer}, {Rudolph}, {Ruiz-Dern}, {Sadowski}, {Sagrist{\`a}
  Sell{\'e}s}, {Sahlmann}, {Salgado}, {Salguero}, {Sarasso}, {Savietto},
  {Schnorhk}, {Schultheis}, {Sciacca}, {Segol}, {Segovia}, {Segransan},
  {Serpell}, {Shih}, {Smareglia}, {Smart}, {Smith}, {Solano}, {Solitro},
  {Sordo}, {Soria Nieto}, {Souchay}, {Spagna}, {Spoto}, {Stampa}, {Steele},
  {Steidelm{\"u}ller}, {Stephenson}, {Stoev}, {Suess}, {S{\"u}veges}, {Surdej},
  {Szabados}, {Szegedi-Elek}, {Tapiador}, {Taris}, {Tauran}, {Taylor},
  {Teixeira}, {Terrett}, {Tingley}, {Trager}, {Turon}, {Ulla}, {Utrilla},
  {Valentini}, {van Elteren}, {Van Hemelryck}, {van Leeuwen}, {Varadi},
  {Vecchiato}, {Veljanoski}, {Via}, {Vicente}, {Vogt}, {Voss}, {Votruba},
  {Voutsinas}, {Walmsley}, {Weiler}, {Weingrill}, {Werner}, {Wevers},
  {Whitehead}, {Wyrzykowski}, {Yoldas}, {{\v{Z}}erjal}, {Zucker}, {Zurbach},
  {Zwitter}, {Alecu}, {Allen}, {Allende Prieto}, {Amorim},
  {Anglada-Escud{\'e}}, {Arsenijevic}, {Azaz}, {Balm}, {Beck}, {Bernstein},
  {Bigot}, {Bijaoui}, {Blasco}, {Bonfigli}, {Bono}, {Boudreault}, {Bressan},
  {Brown}, {Brunet}, {Bunclark}, {Buonanno}, {Butkevich}, {Carret}, {Carrion},
  {Chemin}, {Ch{\'e}reau}, {Corcione}, {Darmigny}, {de Boer}, {de Teodoro}, {de
  Zeeuw}, {Delle Luche}, {Domingues}, {Dubath}, {Fodor}, {Fr{\'e}zouls},
  {Fries}, {Fustes}, {Fyfe}, {Gallardo}, {Gallegos}, {Gardiol}, {Gebran},
  {Gomboc}, {G{\'o}mez}, {Grux}, {Gueguen}, {Heyrovsky}, {Hoar}, {Iannicola},
  {Isasi Parache}, {Janotto}, {Joliet}, {Jonckheere}, {Keil}, {Kim},
  {Klagyivik}, {Klar}, {Knude}, {Kochukhov}, {Kolka}, {Kos}, {Kutka}, {Lainey},
  {LeBouquin}, {Liu}, {Loreggia}, {Makarov}, {Marseille}, {Martayan},
  {Martinez-Rubi}, {Massart}, {Meynadier}, {Mignot}, {Munari}, {Nguyen},
  {Nordlander}, {Ocvirk}, {O'Flaherty}, {Olias Sanz}, {Ortiz}, {Osorio},
  {Oszkiewicz}, {Ouzounis}, {Palmer}, {Park}, {Pasquato}, {Peltzer}, {Peralta},
  {P{\'e}turaud}, {Pieniluoma}, {Pigozzi}, {Poels}, {Prat}, {Prod'homme},
  {Raison}, {Rebordao}, {Risquez}, {Rocca-Volmerange}, {Rosen}, {Ruiz-Fuertes},
  {Russo}, {Sembay}, {Serraller Vizcaino}, {Short}, {Siebert}, {Silva},
  {Sinachopoulos}, {Slezak}, {Soffel}, {Sosnowska}, {Strai{\v{z}}ys}, {ter
  Linden}, {Terrell}, {Theil}, {Tiede}, {Troisi}, {Tsalmantza}, {Tur},
  {Vaccari}, {Vachier}, {Valles}, {Van Hamme}, {Veltz}, {Virtanen}, {Wallut},
  {Wichmann}, {Wilkinson}, {Ziaeepour}, \& {Zschocke}}]{Gaia:2016}
{Gaia Collaboration}, {Prusti}, T., {de Bruijne}, J.~H.~J., {et~al.} 2016,
  \aap, 595, A1

\bibitem[{{Gaia Collaboration} {et~al.}(2023){Gaia Collaboration}, {Vallenari},
  {Brown}, {Prusti}, {de Bruijne}, {Arenou}, {Babusiaux}, {Biermann},
  {Creevey}, {Ducourant}, {Evans}, {Eyer}, {Guerra}, {Hutton}, {Jordi},
  {Klioner}, {Lammers}, {Lindegren}, {Luri}, {Mignard}, {Panem}, {Pourbaix},
  {Randich}, {Sartoretti}, {Soubiran}, {Tanga}, {Walton}, {Bailer-Jones},
  {Bastian}, {Drimmel}, {Jansen}, {Katz}, {Lattanzi}, {van Leeuwen}, {Bakker},
  {Cacciari}, {Casta{\~n}eda}, {De Angeli}, {Fabricius}, {Fouesneau},
  {Fr{\'e}mat}, {Galluccio}, {Guerrier}, {Heiter}, {Masana}, {Messineo},
  {Mowlavi}, {Nicolas}, {Nienartowicz}, {Pailler}, {Panuzzo}, {Riclet}, {Roux},
  {Seabroke}, {Sordo}, {Th{\'e}venin}, {Gracia-Abril}, {Portell}, {Teyssier},
  {Altmann}, {Andrae}, {Audard}, {Bellas-Velidis}, {Benson}, {Berthier},
  {Blomme}, {Burgess}, {Busonero}, {Busso}, {C{\'a}novas}, {Carry}, {Cellino},
  {Cheek}, {Clementini}, {Damerdji}, {Davidson}, {de Teodoro}, {Nu{\~n}ez
  Campos}, {Delchambre}, {Dell'Oro}, {Esquej}, {Fern{\'a}ndez-Hern{\'a}ndez},
  {Fraile}, {Garabato}, {Garc{\'\i}a-Lario}, {Gosset}, {Haigron}, {Halbwachs},
  {Hambly}, {Harrison}, {Hern{\'a}ndez}, {Hestroffer}, {Hodgkin}, {Holl},
  {Jan{\ss}en}, {Jevardat de Fombelle}, {Jordan}, {Krone-Martins}, {Lanzafame},
  {L{\"o}ffler}, {Marchal}, {Marrese}, {Moitinho}, {Muinonen}, {Osborne},
  {Pancino}, {Pauwels}, {Recio-Blanco}, {Reyl{\'e}}, {Riello}, {Rimoldini},
  {Roegiers}, {Rybizki}, {Sarro}, {Siopis}, {Smith}, {Sozzetti}, {Utrilla},
  {van Leeuwen}, {Abbas}, {{\'A}brah{\'a}m}, {Abreu Aramburu}, {Aerts},
  {Aguado}, {Ajaj}, {Aldea-Montero}, {Altavilla}, {{\'A}lvarez}, {Alves},
  {Anders}, {Anderson}, {Anglada Varela}, {Antoja}, {Baines}, {Baker},
  {Balaguer-N{\'u}{\~n}ez}, {Balbinot}, {Balog}, {Barache}, {Barbato},
  {Barros}, {Barstow}, {Bartolom{\'e}}, {Bassilana}, {Bauchet}, {Becciani},
  {Bellazzini}, {Berihuete}, {Bernet}, {Bertone}, {Bianchi}, {Binnenfeld},
  {Blanco-Cuaresma}, {Blazere}, {Boch}, {Bombrun}, {Bossini}, {Bouquillon},
  {Bragaglia}, {Bramante}, {Breedt}, {Bressan}, {Brouillet}, {Brugaletta},
  {Bucciarelli}, {Burlacu}, {Butkevich}, {Buzzi}, {Caffau}, {Cancelliere},
  {Cantat-Gaudin}, {Carballo}, {Carlucci}, {Carnerero}, {Carrasco},
  {Casamiquela}, {Castellani}, {Castro-Ginard}, {Chaoul}, {Charlot}, {Chemin},
  {Chiaramida}, {Chiavassa}, {Chornay}, {Comoretto}, {Contursi}, {Cooper},
  {Cornez}, {Cowell}, {Crifo}, {Cropper}, {Crosta}, {Crowley}, {Dafonte},
  {Dapergolas}, {David}, {David}, {de Laverny}, {De Luise}, {De March}, {De
  Ridder}, {de Souza}, {de Torres}, {del Peloso}, {del Pozo}, {Delbo},
  {Delgado}, {Delisle}, {Demouchy}, {Dharmawardena}, {Di Matteo}, {Diakite},
  {Diener}, {Distefano}, {Dolding}, {Edvardsson}, {Enke}, {Fabre}, {Fabrizio},
  {Faigler}, {Fedorets}, {Fernique}, {Fienga}, {Figueras}, {Fournier},
  {Fouron}, {Fragkoudi}, {Gai}, {Garcia-Gutierrez}, {Garcia-Reinaldos},
  {Garc{\'\i}a-Torres}, {Garofalo}, {Gavel}, {Gavras}, {Gerlach}, {Geyer},
  {Giacobbe}, {Gilmore}, {Girona}, {Giuffrida}, {Gomel}, {Gomez},
  {Gonz{\'a}lez-N{\'u}{\~n}ez}, {Gonz{\'a}lez-Santamar{\'\i}a},
  {Gonz{\'a}lez-Vidal}, {Granvik}, {Guillout}, {Guiraud},
  {Guti{\'e}rrez-S{\'a}nchez}, {Guy}, {Hatzidimitriou}, {Hauser}, {Haywood},
  {Helmer}, {Helmi}, {Sarmiento}, {Hidalgo}, {Hilger}, {H{\l}adczuk}, {Hobbs},
  {Holland}, {Huckle}, {Jardine}, {Jasniewicz}, {Jean-Antoine Piccolo},
  {Jim{\'e}nez-Arranz}, {Jorissen}, {Juaristi Campillo}, {Julbe}, {Karbevska},
  {Kervella}, {Khanna}, {Kontizas}, {Kordopatis}, {Korn}, {K{\'o}sp{\'a}l},
  {Kostrzewa-Rutkowska}, {Kruszy{\'n}ska}, {Kun}, {Laizeau}, {Lambert},
  {Lanza}, {Lasne}, {Le Campion}, {Lebreton}, {Lebzelter}, {Leccia}, {Leclerc},
  {Lecoeur-Taibi}, {Liao}, {Licata}, {Lindstr{\o}m}, {Lister}, {Livanou},
  {Lobel}, {Lorca}, {Loup}, {Madrero Pardo}, {Magdaleno Romeo}, {Managau},
  {Mann}, {Manteiga}, {Marchant}, {Marconi}, {Marcos}, {Marcos Santos},
  {Mar{\'\i}n Pina}, {Marinoni}, {Marocco}, {Marshall}, {Martin Polo},
  {Mart{\'\i}n-Fleitas}, {Marton}, {Mary}, {Masip}, {Massari},
  {Mastrobuono-Battisti}, {Mazeh}, {McMillan}, {Messina}, {Michalik}, {Millar},
  {Mints}, {Molina}, {Molinaro}, {Moln{\'a}r}, {Monari}, {Mongui{\'o}},
  {Montegriffo}, {Montero}, {Mor}, {Mora}, {Morbidelli}, {Morel}, {Morris},
  {Muraveva}, {Murphy}, {Musella}, {Nagy}, {Noval}, {Oca{\~n}a}, {Ogden},
  {Ordenovic}, {Osinde}, {Pagani}, {Pagano}, {Palaversa}, {Palicio},
  {Pallas-Quintela}, {Panahi}, {Payne-Wardenaar}, {Pe{\~n}alosa Esteller},
  {Penttil{\"a}}, {Pichon}, {Piersimoni}, {Pineau}, {Plachy}, {Plum}, {Poggio},
  {Pr{\v{s}}a}, {Pulone}, {Racero}, {Ragaini}, {Rainer}, {Raiteri}, {Rambaux},
  {Ramos}, {Ramos-Lerate}, {Re Fiorentin}, {Regibo}, {Richards}, {Rios Diaz},
  {Ripepi}, {Riva}, {Rix}, {Rixon}, {Robichon}, {Robin}, {Robin}, {Roelens},
  {Rogues}, {Rohrbasser}, {Romero-G{\'o}mez}, {Rowell}, {Royer}, {Ruz Mieres},
  {Rybicki}, {Sadowski}, {S{\'a}ez N{\'u}{\~n}ez}, {Sagrist{\`a} Sell{\'e}s},
  {Sahlmann}, {Salguero}, {Samaras}, {Sanchez Gimenez}, {Sanna},
  {Santove{\~n}a}, {Sarasso}, {Schultheis}, {Sciacca}, {Segol}, {Segovia},
  {S{\'e}gransan}, {Semeux}, {Shahaf}, {Siddiqui}, {Siebert}, {Siltala},
  {Silvelo}, {Slezak}, {Slezak}, {Smart}, {Snaith}, {Solano}, {Solitro},
  {Souami}, {Souchay}, {Spagna}, {Spina}, {Spoto}, {Steele},
  {Steidelm{\"u}ller}, {Stephenson}, {S{\"u}veges}, {Surdej}, {Szabados},
  {Szegedi-Elek}, {Taris}, {Taylor}, {Teixeira}, {Tolomei}, {Tonello}, {Torra},
  {Torra}, {Torralba Elipe}, {Trabucchi}, {Tsounis}, {Turon}, {Ulla}, {Unger},
  {Vaillant}, {van Dillen}, {van Reeven}, {Vanel}, {Vecchiato}, {Viala},
  {Vicente}, {Voutsinas}, {Weiler}, {Wevers}, {Wyrzykowski}, {Yoldas}, {Yvard},
  {Zhao}, {Zorec}, {Zucker}, \& {Zwitter}}]{GAIA:DR3}
{Gaia Collaboration}, {Vallenari}, A., {Brown}, A.~G.~A., {et~al.} 2023, \aap,
  674, A1

\bibitem[{Garavito-Camargo {et~al.}(2019)Garavito-Camargo, Besla, Laporte,
  Johnston, Gómez, \& Watkins}]{Garavito:2019}
Garavito-Camargo, N., Besla, G., Laporte, C. F.~P., {et~al.} 2019, The
  Astrophysical Journal, 884, 51

\bibitem[{{GRAVITY Collaboration} {et~al.}(2019){GRAVITY Collaboration},
  {Abuter}, {Amorim}, {Baub{\"o}ck}, {Berger}, {Bonnet}, {Brandner},
  {Cl{\'e}net}, {Coud{\'e} Du Foresto}, {de Zeeuw}, {Dexter}, {Duvert},
  {Eckart}, {Eisenhauer}, {F{\"o}rster Schreiber}, {Garcia}, {Gao}, {Gendron},
  {Genzel}, {Gerhard}, {Gillessen}, {Habibi}, {Haubois}, {Henning}, {Hippler},
  {Horrobin}, {Jim{\'e}nez-Rosales}, {Jocou}, {Kervella}, {Lacour},
  {Lapeyr{\`e}re}, {Le Bouquin}, {L{\'e}na}, {Ott}, {Paumard}, {Perraut},
  {Perrin}, {Pfuhl}, {Rabien}, {Rodriguez Coira}, {Rousset}, {Scheithauer},
  {Sternberg}, {Straub}, {Straubmeier}, {Sturm}, {Tacconi}, {Vincent}, {von
  Fellenberg}, {Waisberg}, {Widmann}, {Wieprecht}, {Wiezorrek}, {Woillez}, \&
  {Yazici}}]{GRAVITY}
{GRAVITY Collaboration}, {Abuter}, R., {Amorim}, A., {et~al.} 2019, \aap, 625,
  L10

\bibitem[{{Green}(2018)}]{dustmap}
{Green}, G. 2018, The Journal of Open Source Software, 3, 695

\bibitem[{{Harris} {et~al.}(2020){Harris}, {Millman}, {van der Walt},
  {Gommers}, {Virtanen}, {Cournapeau}, {Wieser}, {Taylor}, {Berg}, {Smith},
  {Kern}, {Picus}, {Hoyer}, {van Kerkwijk}, {Brett}, {Haldane}, {del R{\'\i}o},
  {Wiebe}, {Peterson}, {G{\'e}rard-Marchant}, {Sheppard}, {Reddy}, {Weckesser},
  {Abbasi}, {Gohlke}, \& {Oliphant}}]{numpy}
{Harris}, C.~R., {Millman}, K.~J., {van der Walt}, S.~J., {et~al.} 2020, \nat,
  585, 357

\bibitem[{{Harris}(1996{\natexlab{a}})}]{Harris:1996}
{Harris}, W.~E. 1996{\natexlab{a}}, \aj, 112, 1487

\bibitem[{{Harris}(1996{\natexlab{b}})}]{cumulos}
{Harris}, W.~E. 1996{\natexlab{b}}, \aj, 112, 1487

\bibitem[{{Helmi} {et~al.}(2018){Helmi}, {Babusiaux}, {Koppelman}, {Massari},
  {Veljanoski}, \& {Brown}}]{Helmi:2018}
{Helmi}, A., {Babusiaux}, C., {Koppelman}, H.~H., {et~al.} 2018, \nat, 563, 85

\bibitem[{Hui {et~al.}(2017)Hui, Ostriker, Tremaine, \&
  Witten}]{PhysRevD.95.043541}
Hui, L., Ostriker, J.~P., Tremaine, S., \& Witten, E. 2017, Phys. Rev. D, 95,
  043541

\bibitem[{{Hunter}(2007)}]{matplotlib}
{Hunter}, J.~D. 2007, Computing in Science and Engineering, 9, 90

\bibitem[{{Katz} {et~al.}(2023){Katz}, {Sartoretti}, {Guerrier}, {Panuzzo},
  {Seabroke}, {Th{\'e}venin}, {Cropper}, {Benson}, {Blomme}, {Haigron},
  {Marchal}, {Smith}, {Baker}, {Chemin}, {Damerdji}, {David}, {Dolding},
  {Fr{\'e}mat}, {Gosset}, {Jan{\ss}en}, {Jasniewicz}, {Lobel}, {Plum},
  {Samaras}, {Snaith}, {Soubiran}, {Vanel}, {Zwitter}, {Antoja}, {Arenou},
  {Babusiaux}, {Brouillet}, {Caffau}, {Di Matteo}, {Fabre}, {Fabricius},
  {Fragkoudi}, {Haywood}, {Huckle}, {Hottier}, {Lasne}, {Leclerc},
  {Mastrobuono-Battisti}, {Royer}, {Teyssier}, {Zorec}, {Crifo}, {Jean-Antoine
  Piccolo}, {Turon}, \& {Viala}}]{Katz:2023}
{Katz}, D., {Sartoretti}, P., {Guerrier}, A., {et~al.} 2023, \aap, 674, A5

\bibitem[{{Kluyver} {et~al.}(2016){Kluyver}, {Ragan-Kelley}, {P{\'e}rez},
  {Granger}, {Bussonnier}, {Frederic}, {Kelley}, {Hamrick}, {Grout}, {Corlay},
  {Ivanov}, {Avila}, {Abdalla}, {Willing}, \& {Jupyter Development
  Team}}]{jupyter-notebook}
{Kluyver}, T., {Ragan-Kelley}, B., {P{\'e}rez}, F., {et~al.} 2016, in IOS
  Press, 87--90

\bibitem[{Koposov {et~al.}(2023)Koposov, Erkal, Li, Da Costa, Cullinane, Ji,
  Kuehn, Lewis, Pace, Shipp, Zucker, Bland-Hawthorn, Lilleengen, Martell, \&
  Collaboration)}]{koposov:2023}
Koposov, S.~E., Erkal, D., Li, T.~S., {et~al.} 2023, Monthly Notices of the
  Royal Astronomical Society, 521, 4936

\bibitem[{Kruijssen {et~al.}(2020)Kruijssen, Pfeffer, Chevance, Bonaca,
  Trujillo-Gomez, Bastian, Reina-Campos, Crain, \& Hughes}]{Kruijssen:2020}
Kruijssen, J. M.~D., Pfeffer, J.~L., Chevance, M., {et~al.} 2020, Monthly
  Notices of the Royal Astronomical Society, 498, 2472

\bibitem[{{Mart{\'\i}nez-Delgado} {et~al.}(2019){Mart{\'\i}nez-Delgado},
  {Vivas}, {Grebel}, {Gallart}, {Pieres}, {Bell}, {Zivick}, {Lemasle}, {Clifton
  Johnson}, {Carballo-Bello}, {No{\"e}l}, {Cioni}, {Choi}, {Besla}, {Schmidt},
  {Zaritsky}, {Gruendl}, {Seibert}, {Nidever}, {Monteagudo}, {Monelli}, {Hubl},
  {van der Marel}, {Ballesteros}, {Stringfellow}, {Walker}, {Blum}, {Bell},
  {Conn}, {Olsen}, {Martin}, {Chu}, {Inno}, {Boer}, {Kallivayalil}, {De Leo},
  {Beletsky}, {Neyer}, \& {Mu{\~n}oz}}]{Martinez:2019}
{Mart{\'\i}nez-Delgado}, D., {Vivas}, A.~K., {Grebel}, E.~K., {et~al.} 2019,
  \aap, 631, A98

\bibitem[{{Massey} {et~al.}(2010){Massey}, {Kitching}, \&
  {Richard}}]{Massey:2010}
{Massey}, R., {Kitching}, T., \& {Richard}, J. 2010, Reports on Progress in
  Physics, 73, 086901

\bibitem[{{McConnachie}(2012)}]{McConnachie:2012}
{McConnachie}, A.~W. 2012, \aj, 144, 4

\bibitem[{McKinney(2010)}]{pandas}
McKinney, W. 2010, in Proceedings of the 9th Python in Science Conference, ed.
  S.~van~der Walt \& J.~Millman, 51 -- 56

\bibitem[{Mo {et~al.}(2010)Mo, van~den Bosch, \&
  White}]{mo_vandenbosch_white_2010}
Mo, H., van~den Bosch, F., \& White, S. 2010, Galaxy Formation and Evolution
  (Cambridge University Press)

\bibitem[{{Muraveva} {et~al.}(2018){Muraveva}, {Delgado}, {Clementini},
  {Sarro}, \& {Garofalo}}]{Muraveva:2018}
{Muraveva}, T., {Delgado}, H.~E., {Clementini}, G., {Sarro}, L.~M., \&
  {Garofalo}, A. 2018, \mnras, 481, 1195

\bibitem[{Naik \& Widmark(2023)}]{Naik2023-ie}
Naik, A.~P. \& Widmark, A. 2023, Mon. Not. R. Astron. Soc., 527, 11559

\bibitem[{{Paxton} {et~al.}(2011){Paxton}, {Bildsten}, {Dotter}, {Herwig},
  {Lesaffre}, \& {Timmes}}]{MESA_3}
{Paxton}, B., {Bildsten}, L., {Dotter}, A., {et~al.} 2011, \apjs, 192, 3

\bibitem[{{Pe{\~n}arrubia} {et~al.}(2016){Pe{\~n}arrubia}, {G{\'o}mez},
  {Besla}, {Erkal}, \& {Ma}}]{Penarrubia:2015}
{Pe{\~n}arrubia}, J., {G{\'o}mez}, F.~A., {Besla}, G., {Erkal}, D., \& {Ma},
  Y.-Z. 2016, \mnras, 456, L54

\bibitem[{Pedregosa {et~al.}(2011)Pedregosa, Varoquaux, Gramfort, Michel,
  Thirion, Grisel, Blondel, Prettenhofer, Weiss, Dubourg, Vanderplas, Passos,
  Cournapeau, Brucher, Perrot, \& Duchesnay}]{scikit-learn}
Pedregosa, F., Varoquaux, G., Gramfort, A., {et~al.} 2011, Journal of Machine
  Learning Research, 12, 2825

\bibitem[{Perottoni {et~al.}(2022)Perottoni, Limberg, Amarante, Rossi, Queiroz,
  Santucci, Pérez-Villegas, \& Chiappini}]{Perottoni:2022}
Perottoni, H.~D., Limberg, G., Amarante, J. A.~S., {et~al.} 2022, The
  Astrophysical Journal Letters, 936, L2

\bibitem[{{Planck Collaboration} {et~al.}(2020){Planck Collaboration},
  {Aghanim, N.}, {Akrami, Y.}, {Ashdown, M.}, {Aumont, J.}, {Baccigalupi, C.},
  {Ballardi ni, M.}, {Banday, A. J.}, {Barreiro, R. B.}, {Bartolo, N.}, {Basak,
  S.}, {Battye, R.}, {Benabed, K.}, {Bernard, J.-P.}, {B ersanelli, M.},
  {Bielewicz, P.}, {Bock, J. J.}, {Bond, J. R.}, {Borrill, J.}, {Bouchet, F.
  R.}, {Boulanger, F.}, {Bucher, M.}, {Burigana, C.}, {Butler, R. C.},
  {Calabrese, E.}, {Cardoso, J.-F.}, {Carron, J.}, {Challinor, A.}, {Chiang, H.
  C.}, {Chl uba, J.}, {Colombo, L. P. L.}, {Combet, C.}, {Contreras, D.},
  {Crill, B. P.}, {Cuttaia, F.}, {de Bernardis, P.}, {de Zotti, G .},
  {Delabrouille, J.}, {Delouis, J.-M.}, {Di Valentino, E.}, {Diego, J. M.},
  {Dor\'e, O.}, {Douspis, M.}, {Ducout, A.}, { Dupac, X.}, {Dusini, S.},
  {Efstathiou, G.}, {Elsner, F.}, {En\ss{}lin, T. A.}, {Eriksen, H. K.},
  {Fantaye, Y.}, {Farhang, M.}, {Fergusson, J.}, {Fernandez-Cobos, R.},
  {Finelli, F.}, {Forastieri, F.}, {Frailis, M.}, {Fraisse, A. A.},
  {Franceschi, E.}, {Frolov, A.}, {Galeotta, S.}, {Galli, S.}, {Ganga, K.},
  {G\'enova-Santos, R. T.}, {Gerbino, M.}, {Ghosh, T.}, {Gonz\'alez -Nuevo,
  J.}, {G\'orski, K. M.}, {Gratton, S.}, {Gruppuso, A.}, {Gudmundsson, J. E.},
  {Hamann, J.}, {Handley, W.}, {Hansen, F. K.}, {Herranz, D.}, {Hildebrandt, S.
  R.}, {Hivon, E.}, {Huang, Z.}, {Jaffe, A. H.}, {Jones, W. C.}, {Karakci, A.},
  {Keih\ "anen, E.}, {Keskitalo, R.}, {Kiiveri, K.}, {Kim, J.}, {Kisner, T.
  S.}, {Knox, L.}, {Krachmalnicoff, N.}, {Kunz, M.}, {Kur ki-Suonio, H.},
  {Lagache, G.}, {Lamarre, J.-M.}, {Lasenby, A.}, {Lattanzi, M.}, {Lawrence, C.
  R.}, {Le Jeune, M.}, {Lemos, P. }, {Lesgourgues, J.}, {Levrier, F.}, {Lewis,
  A.}, {Liguori, M.}, {Lilje, P. B.}, {Lilley, M.}, {Lindholm, V.}, {L\'opez-Ca
  niego, M.}, {Lubin, P. M.}, {Ma, Y.-Z.}, {Mac\'{\i}as-P\'erez, J. F.},
  {Maggio, G.}, {Maino, D.}, {Mandolesi, N.}, {Mangilli, A.},
  {Marcos-Caballero, A.}, {Maris, M.}, {Martin, P. G.}, {Martinelli, M.},
  {Mart\'{\i}nez-Gonz\'alez, E.}, an~d {Mauri, N.}, {McEwen, J. D.}, {Meinhold,
  P. R.}, {Melchiorri, A.}, {Mennella, A.}, {Migliaccio, M.}, {Millea, M.},
  {Mitra, S.}, {Miville-Desch\^enes, M.-A.}, {Molinari, D.}, {Montier, L.},
  {Morgante, G.}, {Moss, A.}, {Natoli, P.}, {N\o{}rgaard-Niel sen, H. U.},
  {Pagano, L.}, {Paoletti, D.}, {Partridge, B.}, {Patanchon, G.}, {Peiris, H.
  V.}, {Perrotta, F.}, {Pettorino, V.}, {Piacentini, F.}, {Polastri, L.},
  {Polenta, G.}, {Puget, J.-L.}, {Rachen, J. P.}, {Reinecke, M.}, {Remazeilles,
  M.}, {Re nzi, A.}, {Rocha, G.}, {Rosset, C.}, {Roudier, G.},
  {Rubi\~no-Mart\'{\i}n, J. A.}, {Ruiz-Granados, B.}, {Salvati, L.}, {Sandr i,
  M.}, {Savelainen, M.}, {Scott, D.}, {Shellard, E. P. S.}, {Sirignano, C.},
  {Sirri, G.}, {Spencer, L. D.}, an~d {Suur-Uski, A.-S.}, {Tauber, J. A.},
  {Tavagnacco, D.}, {Tenti, M.}, {Toffolatti, L.}, {Tomasi, M.}, {Trombetti,
  T.}, {Valen ziano, L.}, {Valiviita, J.}, {Van Tent, B.}, {Vibert, L.},
  {Vielva, P.}, {Villa, F.}, {Vittorio, N.}, {Wandelt, B. D.}, {W ehus, I. K.},
  {White, M.}, {White, S. D. M.}, {Zacchei, A.}, \& {Zonca, A.}}]{Planck18}
{Planck Collaboration}, {Aghanim, N.}, {Akrami, Y.}, {et~al.} 2020, A\&A, 641,
  A6

\bibitem[{Price-Whelan {et~al.}(2020)Price-Whelan, Sipőcz, Lenz, Greco,
  Starkman, Foreman-Mackey, Lim, Oh, Koposov, \& Major}]{gala2}
Price-Whelan, A., Sipőcz, B., Lenz, D., {et~al.} 2020, adrn/gala: v1.3

\bibitem[{Price-Whelan(2017)}]{gala}
Price-Whelan, A.~M. 2017, The Journal of Open Source Software, 2

\bibitem[{{Riello} {et~al.}(2021){Riello}, {De Angeli}, {Evans}, {Montegriffo},
  {Carrasco}, {Busso}, {Palaversa}, {Burgess}, {Diener}, {Davidson}, {Rowell},
  {Fabricius}, {Jordi}, {Bellazzini}, {Pancino}, {Harrison}, {Cacciari}, {van
  Leeuwen}, {Hambly}, {Hodgkin}, {Osborne}, {Altavilla}, {Barstow}, {Brown},
  {Castellani}, {Cowell}, {De Luise}, {Gilmore}, {Giuffrida}, {Hidalgo},
  {Holland}, {Marinoni}, {Pagani}, {Piersimoni}, {Pulone}, {Ragaini}, {Rainer},
  {Richards}, {Sanna}, {Walton}, {Weiler}, \& {Yoldas}}]{Riello:2021}
{Riello}, M., {De Angeli}, F., {Evans}, D.~W., {et~al.} 2021, \aap, 649, A3

\bibitem[{{Rubin} \& {Ford}(1970)}]{Rubin:1970}
{Rubin}, V.~C. \& {Ford}, W.~Kent, J. 1970, Astrophys. J., 159, 379

\bibitem[{Shipp {et~al.}(2021)Shipp, Erkal, Drlica-Wagner, Li, Pace, Koposov,
  Cullinane, Costa, Ji, Kuehn, Lewis, Mackey, Simpson, Wan, Zucker,
  Bland-Hawthorn, Ferguson, Lilleengen, \& Collaboration)}]{Shipp:2021}
Shipp, N., Erkal, D., Drlica-Wagner, A., {et~al.} 2021, The Astrophysical
  Journal, 923, 149

\bibitem[{{Taylor}(2005)}]{topcat}
{Taylor}, M.~B. 2005, in Astronomical Society of the Pacific Conference Series,
  Vol. 347, Astronomical Data Analysis Software and Systems XIV, ed.
  P.~{Shopbell}, M.~{Britton}, \& R.~{Ebert}, 29

\bibitem[{Tulin \& Yu(2018)}]{TULIN20181}
Tulin, S. \& Yu, H.-B. 2018, Physics Reports, 730, 1, dark matter
  self-interactions and small scale structure

\bibitem[{{van der Marel} {et~al.}(2002){van der Marel}, {Alves}, {Hardy}, \&
  {Suntzeff}}]{vanderMarel:2002}
{van der Marel}, R.~P., {Alves}, D.~R., {Hardy}, E., \& {Suntzeff}, N.~B. 2002,
  \aj, 124, 2639

\bibitem[{{Vasiliev}(2023)}]{Vasiliev:2023}
{Vasiliev}, E. 2023, Galaxies, 11, 59

\bibitem[{{Vasiliev} {et~al.}(2021){Vasiliev}, {Belokurov}, \&
  {Erkal}}]{Vasiliev:2020}
{Vasiliev}, E., {Belokurov}, V., \& {Erkal}, D. 2021, \mnras, 501, 2279

\bibitem[{Villanueva-Domingo {et~al.}(2021)Villanueva-Domingo, Mena, \&
  Palomares-Ruiz}]{10.3389/fspas.2021.681084}
Villanueva-Domingo, P., Mena, O., \& Palomares-Ruiz, S. 2021, Frontiers in
  Astronomy and Space Sciences, 8

\bibitem[{{Virtanen} {et~al.}(2020){Virtanen}, {Gommers}, {Oliphant},
  {Haberland}, {Reddy}, {Cournapeau}, {Burovski}, {Peterson}, {Weckesser},
  {Bright}, {van der Walt}, {Brett}, {Wilson}, {Millman}, {Mayorov}, {Nelson},
  {Jones}, {Kern}, {Larson}, {Carey}, {Polat}, {Feng}, {Moore}, {VanderPlas},
  {Laxalde}, {Perktold}, {Cimrman}, {Henriksen}, {Quintero}, {Harris},
  {Archibald}, {Ribeiro}, {Pedregosa}, {van Mulbregt}, \& {SciPy 1. 0
  Contributors}}]{scipy}
{Virtanen}, P., {Gommers}, R., {Oliphant}, T.~E., {et~al.} 2020, Nature
  Methods, 17, 261

\bibitem[{Wang {et~al.}(2022)Wang, Zhang, Xue, Huang, Liu, Zhang, \&
  Yang}]{wang:2022}
Wang, F., Zhang, H.-W., Xue, X.-X., {et~al.} 2022, Monthly Notices of the Royal
  Astronomical Society, 513, 1958

\bibitem[{Waskom(2021)}]{seaborn}
Waskom, M.~L. 2021, Journal of Open Source Software, 6, 3021

\bibitem[{{Watkins} {et~al.}(2024){Watkins}, {van der Marel}, \&
  {Bennet}}]{Watkins:2024}
{Watkins}, L.~L., {van der Marel}, R.~P., \& {Bennet}, P. 2024, \apj, 963, 84

\bibitem[{{Weinberg}(1986)}]{1986ApJ...300...93W}
{Weinberg}, M.~D. 1986, Astrophysical Journal, 300, 93

\bibitem[{Zavala \& Frenk(2019)}]{galaxies7040081}
Zavala, J. \& Frenk, C.~S. 2019, Galaxies, 7

\bibitem[{{Zonca} {et~al.}(2019){Zonca}, {Singer}, {Lenz}, {Reinecke},
  {Rosset}, {Hivon}, \& {Gorski}}]{healpy}
{Zonca}, A., {Singer}, L., {Lenz}, D., {et~al.} 2019, The Journal of Open
  Source Software, 4, 1298

\bibitem[{Zwicky(1933)}]{Zwicky:1933}
Zwicky, F. 1933, Helv. Phys. Acta, 6, 110

\end{thebibliography}

\begin{appendix}
    \section{Large Magellanic Cloud rest frame}
    \label{apendice}
    \subsection{Centre-of-mass rest frame}
    \label{apendice-cm}
The coordinate system used to compare the data with the theoretical model has its origin in the LMC and SMC CM and the $x$ axis orientated according to the DM subhalo velocity, $\bar{v}_s$. In order to obtain the coordinates of our dataset in such a rest frame, we performed the following steps:
\begin{enumerate}
\item we boosted the data to the new frame by ${\bar{r}_{boost}=\bar{r}_{obs}-\bar{r}_{cm}}$;
\item we performed the rotations upon the boosted data using the matrix
\end{enumerate}
\begin{align}
M&=\left(
\begin{array}{ccc}
\cos \theta_1 \cos \theta_2 & \sin \theta_1 \cos \theta_2 & \sin \theta_2 \\
-\sin \theta_1 &\cos \theta_1 &0 \\
- \cos \theta_1 \sin \theta_2 & -\sin \theta_1 \sin \theta_2 &\cos \theta_2 
\end{array}
\right)\,,
\end{align}
to obtain the coordinate $\left(x',\, y',\, z'\right)$ in the new rest frame. The velocity had to be transformed as well and the final velocity had to be boosted using $\bar{v}_s=v_s \hat{i}'$. The angles are defined as
\begin{align}
\tan \theta_1&=\frac{\left(v_{cm}\right)_y}{\left(v_{cm}\right)_x} \,,\\
\tan \theta_2&=\frac{\left(v_{cm}\right)_z}{\sqrt{\left(v_{cm}\right)_x^2+\left(v_{cm}\right)_y^2}}\,,
\end{align}
where $\left(\left(v_{cm}\right)_x, \, \left(v_{cm}\right)_y, \,  \left(v_{cm}\right)_z\right)$ is the CM velocity in the solar coordinate system.

\subsection{Orbit frame}
\label{apendice:orbit}
The coordinate system used to compare our results with the Fig. 1 presented by \citet{Buschmann:2018} has its origin in the CM of the MCS, its $z^*$ axis perpendicular to the orbital plane, and the $x^*$ axis orientated in the direction of the velocity of the DM subhalo. To obtain these new coordinates from the CM rest frame, we had to perform a rotation of an angle, $\theta_{orb}$, according to the following matrix:
\begin{eqnarray}
M_{orb}&=&\left(
\begin{array}{ccc}
1 & 0 & 0 \\
0 &\cos \theta_{orb} &\sin \theta_{orb} \\
0 & -\sin \theta_{orb}  &\cos \theta_{orb} 
\end{array}
\right)\,,
\end{eqnarray}
where $\tan \theta_{orb}=\overline{z}_{orb}/\overline{y}_{orb}$, $\overline{z}_{orb}$ $\left(\overline{y}_{orb}\right)$ is the mean value of the $z$ $\left(y\right)$ coordinate in the CM rest frame of the CM's past orbit.

\section{Estimation of radial velocities using machine learning}
\label{apendice:machinelearning}

To complete the phase information for all the halo stars in our sample, we applied two machine learning techniques to assign radial velocities to those stars without such measurements.
\subsection{K-giant radial velocity}

Our first sample consists of 245086 K-giant stars, of which 10989 have measured radial velocities. The last subsample was used to train a random forest regressor (RF) \citep{Breiman:2001}. The chosen predictors for this study were the angular coordinates, proper motions, G magnitudes, BP and RP colours, distances to the Sun and the galactic centre, and the Galactocentric Cartesian coordinates. To prevent overfitting, a standard cross-validation analysis was performed. The RF hyper-parameters were tuned using GridSearchCV from the Scikit Learn library, resulting in the following values: {[$n\_estimators=4900$, $random\_state = 0$]}
 
\subsection{RR-Lyrae radial velocity}

We observed a decrease in the number of measured radial velocities for RR-Lyrae stars, resulting in a drop in the fraction of measured radial velocities to (5510/67276), considering galactocentric distances between 10 to 100 kpc. To tackle this issue, we aimed to model the spatial distribution of radial velocities. To achieve this, we utilised normalising flows (NFs) \citep{durkan2019neural}, as have been implemented by \citet{2021AAS...23823001C}, to model the joint posterior probability of radial velocities and predictors using a subsample of the features described earlier. However, we limited the feature space to prevent any bias against less luminous stars. Therefore, we did not consider the magnitude and colours of stars as predictors in this case. 

We used probabilistic modelling to generate a radial velocity distribution outcome for a significant number of stars (100000). Therefore, the normalizing flow was used to augment and generalise the training dataset. We evaluated the marginal probability of the radial velocity given the values of other variables (predictors) and obtained a vector of probabilities. To complement this method, we used the RF algorithm to map the posterior conditional radial velocity distribution to the real measured value in the training sample, similar to the method used for K-giant stars. The RF hyper-parameters were tuned using GridSearchCV from the Scikit Learn library, resulting in the following values:
[$max\_depth=50$, $max\_features=8$, $min\_samples\_leaf=1$, $min\_samples\_split=6$, $n\_estimators=800$]. It is important to note that the result of the NFs is a vector that represents the conditional radial velocity. The vector is measured on a 1000-dimensional grid that samples velocities ranging from -700 to 700 km/s. To prevent overfitting, cross-validation was employed by splitting the data in two and using 80$\%$ for training and 20$\%$ for validation, similar to the previous case.\\

\subsection{Goodness of fit in radial velocity regression}

R-Squared ($R^2$), or the coefficient of determination, is a statistical measure used to determine the proportion of variance in the dependent variable that can be explained by the independent variable in a regression model. The statistics were calculated for our two samples of stars, resulting in values of 0.86 for K-giant stars and 0.5 for RR-Lyrae ones, respectively. These values are comparable to those recently reported by \cite{Naik2023-ie} using Bayesian neural networks, and provide us with a complete sample of halo stars in the phase space.

We tested the inferred stars' radial velocities by measuring the mean radial velocities of satellite galaxies and compared them with the tabulated velocity values \citep{McConnachie:2012}. The results are presented in Table. \ref{vel}. Our machine-learning results show that the average velocity of the stars on these satellite galaxies is well reproduced. The relevant error of the average velocity for the Sculptor galaxy is due to the lack of any measured radial velocity on this object.

\begin{table}[h!]
    \centering
    \begin{tabular}{c c c}
    \hline\hline
        Galaxy & $V_r$ tabulated [km/s]& $V_r$ inferred [km/s]  \\
        \hline
      LMC   &262.2&250.62$\pm$30.40\\
      SMC&145.6&133.64$\pm$30.51\\
      Carina&222.9&213.00$\pm$41.37\\
      Draco&-291.0&-186.10$\pm$45.91\\
      Sculptor &111.4&-0.88$\pm$149.37\\
      \hline
    \end{tabular}
    \caption{Comparison between the mean radial velocity inferred with machine learning and the values tabulated for MW satellite galaxies \citep{McConnachie:2012}.}
    \label{vel}
\end{table}

Using this method, we can estimate the mean velocity field in the halo, which is necessary for the likelihood estimation of the subhalo mass of the LMC.

\section{Validation of estimated distances for K-giant stars}

\label{apendice:distancias}
To test the photometric distance obtained for the K giants, we computed the photometric distances of the globular clusters NGC7006, NGC5694, NGC2419, and NGC6229, and of the LMC, SMC, Carina, Draco, and Sculptor. 

The observational data was extracted from \textit{Gaia} Data Release 3 \citep{GAIA:DR3,Gaia:2016}, using the option single object searcher to avoid field-contamination. Additionally, the data for the LMC and the SMC were selected on a circular disk in $(l,b)$ centred on the location of each MC, with a radius of $2^{\circ}$. For Carina, Draco, and Sculptor we selected data using an angular mask of three times the tidal radius reported in \citet{Drlica-Wagner:2020}. We performed the data reduction or treatment indicated in the text for the K giant (that is, corrected for dust extinction, discarded all sources with $E(B-V) > 0.3$, corrected the magnitudes, and performed the $3\sigma$ cut upon the corrected BP and RP flux excess factor ($C^*$)). Finally, we selected the giant branch and performed the cross-match with the K-giant catalogue given by \textit{Gaia} (gaiadr3.astrophysical parameters). For the LMC and SMC, we also restricted the dataset in the reported \textit{Gaia} parallax.

\begin{figure}[h!]
\centering
\includegraphics[width=\hsize]{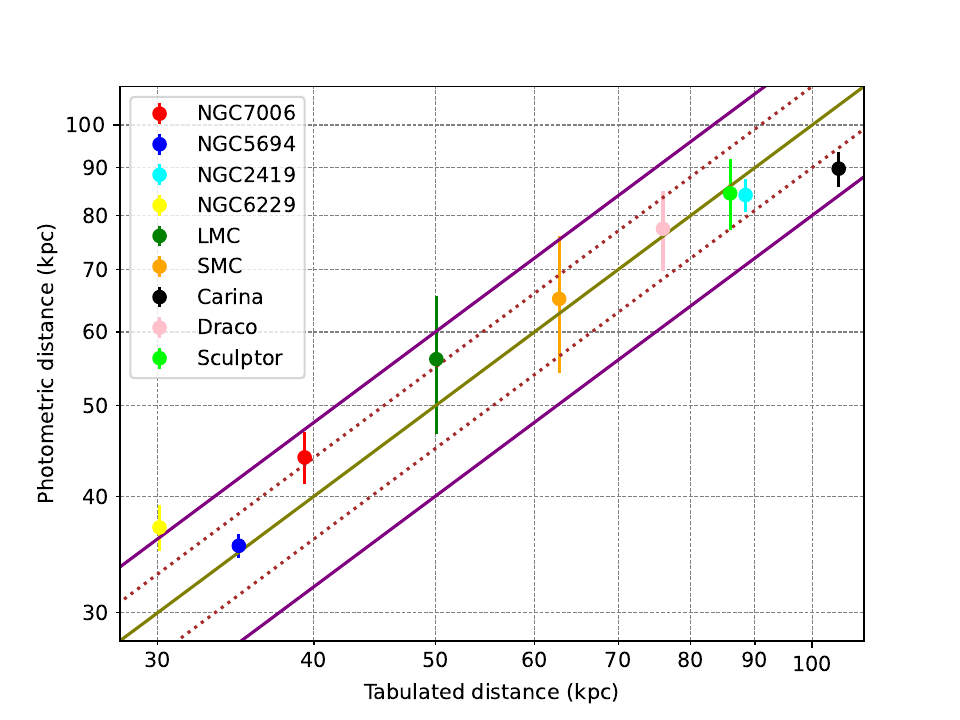}
\caption{ Comparison between the photometric distances computed using Eq. (\ref{mg}) and the tabulated distances \citep{cumulos,galaxias}. Solid olive line: photometric distance equals tabulated distance. Dashed brown line: $\pm 10\%$ with respect to the one-to-one line. Solid purple line: $\pm 20\%$ with respect to the one-to-one line. The vertical lines are the statistical errors.}
\label{dist_fig}
\end{figure}

We calculated the photometric distance of each source and the mean value of each object. The results are shown in Fig. \ref{dist_fig} and, as can be seen, the computed photometric distances and the values reported in the literature \citep{cumulos,galaxias} are in good agreement.

We have also applied our fit to K-giant stars from the \citet{Conroy:2021} catalogue and found that our calculated distances are in agreement with the distances reported by \citet{Conroy:2021} within 10$\%$.

\end{appendix}

\end{document}